# A Taxonomic Approach to Understanding Emerging Blockchain Identity Management Systems

Loïc Lesavre
Priam Varin
Peter Mell
Michael Davidson
James Shook
*Computer Security Division*
*Information Technology Laboratory*

January 14, 2020

**NIST**
**National Institute of**
**Standards and Technology**
U.S. Department of Commerce

## Abstract

Identity management systems (IDMSs) are widely used to provision user identities while managing authentication, authorization, and data sharing within organizations and on the web. Traditional identity systems typically suffer from single points of failure, lack of interoperability, and privacy issues, such as enabling mass data collection and user tracking. Blockchain technology has the potential to alleviate these concerns: it can support the ability for users to control the custody of their own identifiers and credentials, enabling novel data ownership and governance models with built-in control and consent mechanisms. Hence, blockchain-based IDMSs, which could benefit both users and businesses, are beginning to proliferate. This work categorizes these systems into a taxonomy based on differences in blockchain architectures, governance models, and other salient features. Context is provided for the taxonomy through the description of related terms, emerging standards, and use cases while highlighting relevant security and privacy considerations.

## Disclaimer

Any mention of commercial products or reference to commercial organizations is for information only; it does not imply recommendation or endorsement by the National Institute of Standards and Technology (NIST), nor does it imply that the products mentioned are necessarily the best available for the purpose.

## Additional Information

For additional information on NIST's Cybersecurity programs, projects and publications, visit the Computer Security Resource Center. Information on other efforts at NIST and in the Information Technology Laboratory (ITL) is also available.

## Comments on this publication may be submitted to:

National Institute of Standards and Technology
Attn: Computer Security Division, Information Technology Laboratory
100 Bureau Drive (Mail Stop 8930) Gaithersburg, MD 20899-8930
Email: blockchain-idms-paper@nist.gov

All comments are subject to release under the Freedom of Information Act (FOIA).

## Acknowledgments

The authors wish to thank all contributors to this publication and their colleagues who reviewed drafts of this report, contributed technical and editorial additions, and supported this work. This includes NIST staff and associates Dylan Yaga, Frederic De Vaulx, Angela Robinson, and Katya Delak.

Additional thanks to all the people and organizations who submitted comments during the public comment period.

## Audience

This publication is designed for readers with some knowledge of blockchain technology who wish to understand how blockchain identity management systems work at a high level, what they offer, and how to better distinguish the emerging building blocks and architectures. It is not intended to be a technical guide; the discussion of the technology provides a conceptual understanding. Note that some examples, figures, and tables are simplified to fit the audience.

## Executive Summary

Identity management systems allow one to provision identities to users while managing authentication, authorization, and data sharing within organizations and on the web. With traditional identity management, organizations store the credentials (e.g., a password) of each user they interact with, and with federated models, they use a third party to store this information. This creates interoperability, security, and privacy concerns, such as data leaks, due to the privileged position of the entity that controls the identity information.

A possible solution to these issues is found in the use of blockchain technologies for identity management: they can support the ability for users to control the custody of their own identifiers and credentials, transform data governance models, reduce dependence on trusted intermediaries, and benefit both users and businesses. Users can manage their identity data themselves and disclose it directly to relying parties on a need-to-know basis (through self-custody of their identifiers and credentials or designating third-party custodians). Businesses can streamline their operations by relying on verifiable user information without having to act as data custodians themselves and dealing with the associated costs and risks (e.g., for infrastructure, security, and regulatory compliance).

A large number of blockchain-based identity management approaches are currently being explored, implemented, and commercialized. Many of them use or plan to use smart contracts, the privacy capabilities gained from zero-knowledge proofs, and scalability solutions atop the underlying blockchain. This is an emerging field, and the features, capabilities, security, and privacy of these proposed systems need to be carefully scrutinized.

The systems that are being designed take architectural forms with varying uses of blockchain and follow types of governance models spanning from top-down approaches to self-sovereign bottom-up ones. Each system has different control and delegation capabilities as well as scalability constraints.

This work breaks down identifier and credential architectures, discusses their reliance on blockchains, and possible combination patterns. It looks at the levels at which onchain registries are created and who can control them. It investigates bring-your-own blockchain address schemes along with credentials being issued as offchain objects. It does not attempt to judge between the different architectures and models but instead highlights their differences.

This document first offers a terminology, list of standards, and fundamental building blocks of blockchain-based identity management. It then provides a breakdown of distinguishing properties and architectures. Next, it discusses public registries and system governance. Finally, it covers some of the security concerns that can affect these systems, as well as additional considerations on core blockchain protocols, zero-knowledge proofs, presentation sharing, data mining, and use cases.

This paper aims to help the reader navigate how blockchain-based identity management systems work, what they offer, and how to better distinguish the emerging building blocks and architectures.

## Table of Contents

## List of Appendices

## 1    Introduction

A large number of blockchain-based identity management approaches are being explored, implemented, and commercialized. This is a new field, and the features, security, and privacy of these proposed systems are often unclear. While many of the approaches hold great promise, most projects rely on the prerequisite of using a blockchain platform that is reliable, secure, scalable, and, sometimes, publicly accessible. Thus, blockchain-based identity management systems are an emerging area that should be watched and carefully evaluated as a potential but not guaranteed breakthrough for digital identity and data ownership.

### 1.1    Background

Identity management systems (IDMSs) are a foundational infrastructure for interactions between entities (individuals, organizations, or things) to support commerce, education, health care, government services, and many other aspects of society. An IDMS must allow entities to authenticate while at the same time distributing information about those entities to enable the granting of access privileges of differing levels or types.

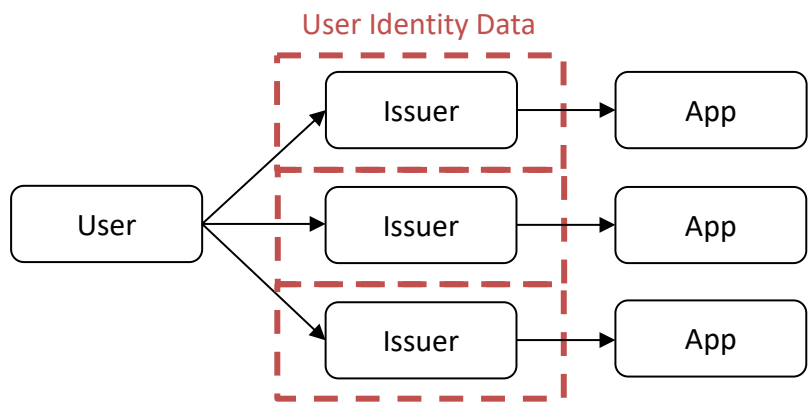

**Figure 1: Traditional Identity Management**

With traditional identity management, businesses store credentials about each user with which they interact (e.g., a password). This enables a user to directly authenticate to the business (i.e., the *relying party*) with which they need to interact, as shown in Figure 1. However, the user is burdened with needing to separately authenticate to each business using different credentials. In addition, businesses are unable to automatically obtain and evaluate verifiable identifying information about each user.

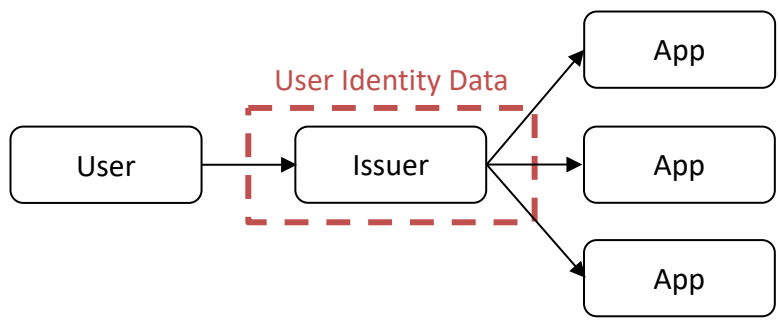

**Figure 2: Federated Identity Management**

More recently, federated identity management systems [1] enable credential service providers to maintain user credentials on behalf of various relying parties. This supports *single sign-on* capabilities where a user utilizes a single set of credentials to access a large number of services, as shown in Figure 2. This frees up the user from having to maintain many passwords. However, it creates interoperability, security, and privacy concerns given the privileged position of the credential service provider between the user and relying parties. For example, it presents a single point of failure that could inhibit the user's access to relying parties or, even worse, that could enable the credential service provider to masquerade as a user.

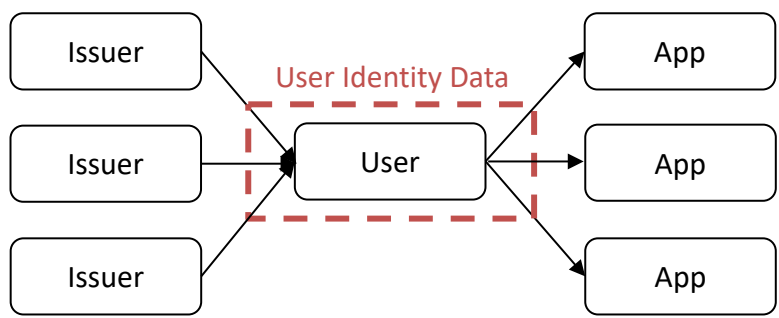

**Figure 3: User-Centric Identity Management**

Blockchain is a technology that could help address these issues by making it easier for users to control the custody of their identity data and interact with relying parties directly instead of relying on a trusted intermediary. As shown in Figure 3, this architecture follows a *user-centric model* of identity management—users control custody through either exclusive *self-custody* of their identifiers and credentials or by choosing third-party custodians to do so on their behalf. This puts users in a position where they can share verifiable information on a need-to-know basis.

From the relying party's perspective, blockchain-based IDMSs allow them to verify that some user information needed to initiate a transaction is valid without having to act as a custodian and store user profile and identity data themselves. A key implication of reduced data custody is that it can lower costs (e.g., for infrastructure, security, and regulatory compliance), privacy and security burdens, and barriers to bootstrapping new business activities.

Note that this architecture could facilitate the development of applications that do not store user identity and session data. User permission may then be necessary for an application to access user profile and session data prior to executing a given transaction (e.g., opening a bank account).

In summary, blockchain-based IDMSs have the potential to enhance security and privacy and grant built-in control and consent capabilities for both users and relying parties. However, there are tradeoffs to be made, and it is necessary to carefully evaluate those emerging solutions.

## 1.2   Purpose and Scope

This document provides an introduction to the different blockchain identity management approaches currently being explored and implemented. The purpose is not to review each solution individually but to provide a taxonomic approach that gives the reader different viewpoints and methods by which blockchain-based identity management can be designed. In this way, the document highlights the different features and characteristics that are possible while exploring the opportunities, challenges, and risks they represent.

As an emerging field, weaknesses may become evident that negate otherwise apparent advantages, or other data models may emerge with even greater benefit. Blockchain infrastructures, identity management platforms, and related user tools are still maturing, with efforts being developed to improve scalability and reliability. While the time for most readers to deploy these capabilities lies somewhere in the future, the capabilities and architecture designs discussed in this paper represent a potential new paradigm for digital identity, which could remodel or even replace traditional and federated identity management systems. Therefore, this field deserves careful consideration and scrutiny today, and this paper is intended to provide the foundational tools to enable such an ongoing evaluation.

## 1.3   Disclaimers and Clarifications

Although *blockchains*, *smart contracts*, and related concepts and technologies are referred to and examined throughout the paper, no recommendation or endorsement of any particular protocol is provided, and the work may be extended to other types of distributed ledger technologies (DLT). Any products or protocols mentioned are for explanatory purposes only and do not imply any endorsement or suitability for use. Regulatory compliance is also out of scope.

## 1.4   Blockchain Identity Management Initiatives and Guidance

Some organizations have already written guidance on blockchain in identity management. The European Union recently published *Blockchain for Government and Public Services* [2] and *Blockchain and Digital Identity* [3]. In the United States, there have been initiatives led at the state level, such as the Illinois Blockchain Initiative [4]. The American Council for Technology and Industry Advisory Council (ACT-IAC) published a *Blockchain Primer* [5] to introduce how blockchain could impact the U.S. Federal Government as well as a *Blockchain Playbook* [6] to introduce how it could be applied to the Federal Government for different purposes, including identity management.

There are a handful of blockchain-based IDMS pilots as well. Some organizations have already adopted the use case of diploma and certificate issuance on the blockchain, such as the Massachusetts Institute of Technology with Blockcerts and Learning Machine [7]. Some jurisdictions are experimenting with blockchain-based IDMSs at different levels, such as Estonia (for electronic medical records) [8], the City of Zug in Switzerland using uPort (on the Ethereum blockchain) [9], and the Provinces of British Columbia and Ontario in Canada using the Verifiable Organizations Network (VON) on the Sovrin blockchain [10].

Many projects are currently under active development, characterizing the growing interest in blockchain-based identity management.

## 1.5    Results of the Public Comment Period

This document has seen minor revisions in response to the comments received during the public comment period that took place from July 9, 2019 to August 9, 2019. A new section (Section 4.4.1.3, *Unspent Transaction Output Model*), a number of subsections (*DID Auth* in Section 3.3, *User Profile Data Management Protocols* in Section 3.4, and *Verifying Business Identity* in Section 7), and figures (Figures 1, 2, and 3) have been added. Other clarifications, add-ons, and/or corrections were incorporated throughout the document. All the sections present in the draft were kept.

## 1.6    Document Structure

The rest of this document is organized as follows:

- **Section 2** introduces blockchain technology and smart contracts at a high-level.
- **Section 3** defines terminology and discusses emerging standards and building blocks for blockchain identity management.
- **Section 4** introduces and discusses a taxonomy in the form of distinguishing properties, which are then used to characterize different architecture designs.
- **Section 5** introduces some security concerns and their mitigation mechanisms.
- **Section 6** provides additional considerations.
- **Section 7** introduces potential use cases.
- **Section 8** is the conclusion.
- **References** lists the references used throughout the document.
- **Appendix A** provides a list of acronyms and abbreviations used in the document.
- **Appendix B** contains a glossary for selected terms defined in the document.

## 2    Blockchains and Smart Contracts

Readers who have little or no knowledge of blockchain technology and who wish to understand how it works at a high level are invited to read National Institute of Standards and Technology Internal Report (NISTIR) 8202, *Blockchain Technology Overview* [11]. It defines blockchain as "tamper evident and tamper resistant digital ledgers implemented in a distributed fashion (i.e., without a central repository) and usually without a central authority (i.e., a bank, company or government). At their basic level, they enable a community of users to record transactions in a shared ledger within that community, such that under normal operation of the blockchain network no transaction can be changed once published."

The technology is called blockchain because the transactions are grouped and published separately in distinct data structures called blocks, which are cryptographically linked together and distributed in a peer-to-peer network to prevent tampering of previously published transactions. Blockchain accounts are based on asymmetric-key cryptography and allow participants to sign transactions. The transactions are added to blocks that must be validated by the nodes that are running the blockchain's peer-to-peer node client. Consensus models determine which node gets the privilege of publishing the next block (see Section 4.6 on *Consensus Comparison Matrix* of NISTIR 8202).

The paper discusses two important categories that pertain to our investigation of identity management systems: "Blockchain networks can be categorized based on their permission model, which determines who can maintain them (e.g., publish blocks). If anyone can publish a new block, it is *permissionless*. If only particular users can publish blocks, it is *permissioned*. In simple terms, a permissioned blockchain network is like a corporate intranet that is controlled, while a permissionless blockchain network is like the public internet, where anyone can participate. Permissioned blockchain networks are often deployed for a group of organizations and individuals, typically referred to as a consortium."

Some blockchains have highly expressive, native smart contract capabilities which can be useful for blockchain-based identity management solutions. A smart contract is defined as: "a collection of code and data (sometimes referred to as functions and states) that is deployed using cryptographically signed transactions on the blockchain network (e.g., Ethereum's smart contracts, Hyperledger Fabric's chaincode). The smart contract is executed by nodes within the blockchain network; all nodes that execute the smart contract must derive the same results from the execution, and the results of execution are recorded on the blockchain. […] The smart contract code can represent a multi-party transaction, typically in the context of a business process. In a multi-party scenario, the benefit is that this can provide attestable data and transparency that can foster trust, provide insight that can enable better business decisions, reduce costs from reconciliation that exists in traditional business to business applications, and reduce the time to complete a transaction. […] Smart contracts must be deterministic, in that given an input, they will always produce the same output based on that input." Furthermore, a source of data from outside a blockchain that serves as input for a smart contract is referred to as an *oracle*.

The owner of a blockchain identity management system does not necessarily own the blockchain upon which the system is built. In fact, an entity can deploy an identity management system without having to build or maintain the underlying blockchain infrastructure that is being used, provided that this meets the entity's needs and risk tolerance.

## 3       Fundamentals of Blockchain Identity Management

Prior to the introduction of the taxonomy in the next section, this section details key terminology, common roles and objects, emerging supportive standards, essential building blocks, and a blockchain identity management communication stack. These terms, standards, and abstractions are used by most blockchain-based identity management systems.

### 3.1    Terminology

Specialized terminology is used for blockchain-based identity management, though it inherits some domain-specific terms related to identity management in general. Understanding the following terms is necessary in order to understand the concepts discussed in this paper:

**Claim**: A characteristic or statement about a *subject* made by an *issuer* as part of a *credential*.

**Credential**: A set of one or more *claims* made by an *issuer*. A *credential* is associated with an *identifier*.

**Custodian**: An *entity* acting on behalf of another *entity* with respect to their *identifiers* and/or *credentials*.

**Entity**: A person, organization, or thing.

**Holder**: A *custodian* holding a *credential* on behalf of an *entity*.

**Identifier**: A blockchain address or other pseudonym that is associated with an *entity*.

**Issuer**: An *entity* that issues a *credential* about a *subject* on behalf of a *requester.*

**Presentation**: Information derived from one or more *credentials* that a *subject* discloses to a *verifier* to communicate some quality about a *subject*.

**Relying Party**: An *entity* that receives information about a subject from a *verifier.*

**Requester**: An *entity* that makes a request to an *issuer* to issue a *credential* about a *subject*.

**Subject**: An *entity* that receives one or more *credentials* from an *issuer*.

**System Owner**: An *entity* that owns a given identity management system.

**Verifier**: An *entity* that verifies the validity of a *presentation* on behalf of a *relying party*.

## 3.2   Blockchain Identity Management Roles and Object Relationships

With this terminology, the common roles that occur in blockchain-based IDMSs and the relationships between these roles can be identified, as well as common objects found in these systems and the relationships between those objects.

Figure 4 provides a high-level overview of the identity management roles defined in the previous section as follows:

- *Requesters*, *issuers*, *subjects* and *holders* are involved in credential issuance.
- *Subjects*, *holders*, *verifiers*, and *relying parties* are involved in presentation disclosure.
- *Requesters* ask for the issuance of a credential from *issuers*.
- *Issuers* provide credentials to *subjects* (or *holders*).
- *Subjects* (or *holders*) reveal presentations to *verifiers.*
- *Verifiers* verify presentations on behalf of *relying parties*.

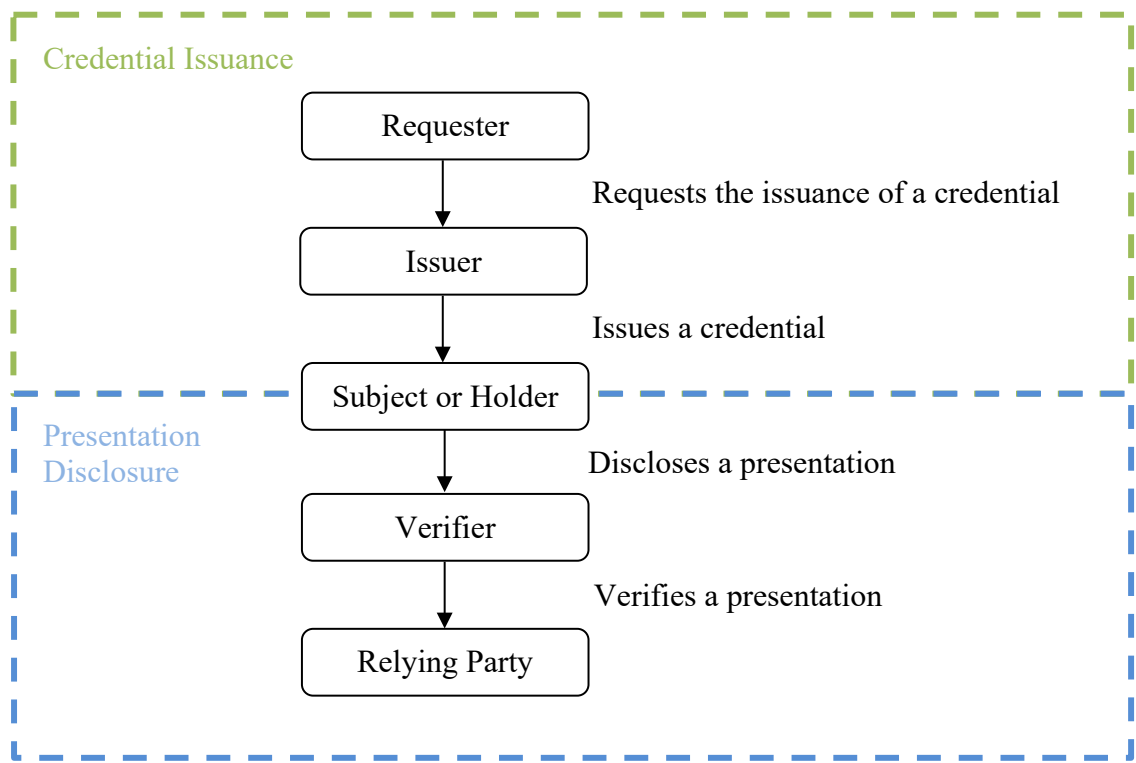

**Figure 4: Identity Management Roles**

Note that these roles are not exclusive. For instance, a *subject* and an *issuer* can both take the *requester* role; a *subject* and a *verifier* can both be a *relying party*. Depending on the IDMS, the approval of a *subject* may be required to issue a new credential to that *subject*.

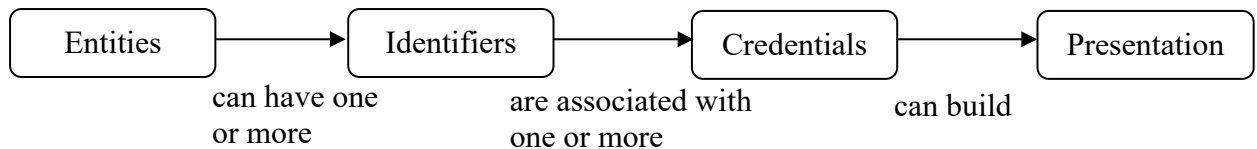

**Figure 5: Hierarchy of IDMS Objects**

Figure 5 provides a high-level overview of the objects that entities interact with in a blockchain-based IDMS. The figure shows that entities can have one or more identifiers, that identifiers are associated with one or more credentials, and that presentations are derived from credentials.

## 3.3  Emerging Standards

There is a set of emerging standards that support novel forms of IDMSs—and blockchain-based ones in particular—using features and properties provided by blockchain, cryptography, and other related technologies. They include:

- Decentralized Identifiers (DIDs) and Verifiable Credentials from the World Wide Web Consortia (W3C)
- Universal Resolver and Identity Hubs from the Decentralized Identity Foundation (DIF)
- DID Auth from the Rebooting the Web-of-Trust (RWOT) working group
- Open Badges from Mozilla and IMS Global

In the subsequent sections of this paper, the terms identifiers, credentials, and presentations will be used but will not necessarily refer to standards of this section.

In addition, blockchain network-specific standards, such as Ethereum Request for Comments (ERCs) and Bitcoin Improvement Proposals (BIPs), will be referred to.

**Decentralized Identifiers – W3C**:

Decentralized Identifiers (DIDs) [12] are identifiers whose purpose is to facilitate the creation of persistent encrypted private channels between entities without the need for any central registration mechanism. They can be used, for example, for credential exchanges and authentication.

An entity can have multiple DIDs, even one or more per relationship with another entity (see *Pairwise-Pseudonymous and Single-Use Identifiers* in Section 4.3). Ownership of a DID is established by demonstrating possession of the private key associated with the public key bound to the DID.

A DID method is a public, standard set of schemes by which to create, resolve, update, and delete DIDs. These methods allow for DID registration, replacement, rotation, recovery, and expiration within an IDMS.

A DID has the following format:

"did:" + <did-method> + ":" + <method-specific-identifier>

As an example, a DID for a "NIST DID method" could look like: *did:nist:0x1234abcd*.

As part of a DID method, a DID resolver allows one to take a DID as input and return the associated metadata, called DID document, which follows formats such as JavaScript Object Notation (JSON) and its related JavaScript Object Notation for Linked Data (JSON-LD) [13] (JSON-LD is a JSON-based format used to serialize linked data).

According to W3C's primer [14], a DID document is comprised of the following standard elements:

- Uniform Resource Identifier (URI) to identify terminology and protocols that allow parties to read the DID document
- DID that identifies the subject of the DID document
- Set of public keys used for authentication, authorization, and communication mechanisms
- Set of authentication methods used for the DID subject to prove ownership of the DID to another entity
- Set of authorization and delegation methods for the DID subject to allow another entity to operate on their behalf (i.e., custodians)
- Set of service endpoints to describe where and how to interact with the DID subject
- Timestamp for when the document was created [optional]
- Timestamp for when the document was last updated [optional]
- Cryptographic proof of integrity (e.g., digital signature) [optional]

The DIF is developing a related specification, called Well-Known DID Configuration [15], which enables the linkage between DIDs and web domains. An entity that controls both a DID and a web domain can prove that they are the same entity.

**Verifiable Credentials and Verifiable Presentations – W3C:**

The Verifiable Credentials specification [16] defines a format for credentials that can be exchanged between DIDs. A Verifiable Credential is a tamper-resistant credential that is cryptographically signed by its issuer.

A Verifiable Credential commonly includes:

- URI to identify the credential and/or the subject of the credential (e.g., DID) [optional]
- URI to identify the credential type
- Claims data or metadata to access it
- URI to identify the issuer (e.g., DID)
- Issuance date
- Cryptographic proof of the issuer (e.g., digital signature)
- URI to identify terminology and protocols that allow parties to read the credential
- Expiration conditions [optional]
- Location of credential status [optional]

The W3C specification also defines Verifiable Presentations. A Verifiable Presentation is a tamper-resistant presentation derived from one or more Verifiable Credentials and cryptographically signed by the subject disclosing it.

A Verifiable Presentation is typically composed of the following:

- URI to uniquely identify the presentation
- URI to identify the type of the object
- One or more verifiable credentials or data derived from them
- URI to identify the entity generating the presentation (e.g., DID)
- Cryptographic proof of the subject (e.g., digital signature)

**Universal Resolver – DIF**:

While DID documents can be retrieved through using the associated DID resolvers, there are advantages to having a more general resolver that can communicate with multiple identifier management systems (including DID systems). The Universal Resolver [17] achieves this goal by enabling application code to be written for a single resolver interface that integrates with multiple identifier management systems.

A DID-based blockchain IDMS that supports the Universal Resolver must define and implement a DID driver that links the Universal Resolver to their system-specific DID method for reading DID documents. This allows applications relying on the IDMS to query DIDs in a common interface so that they do not have to deal with fetching the system-specific DID methods themselves. This takes place according to the steps shown in Figure 6.

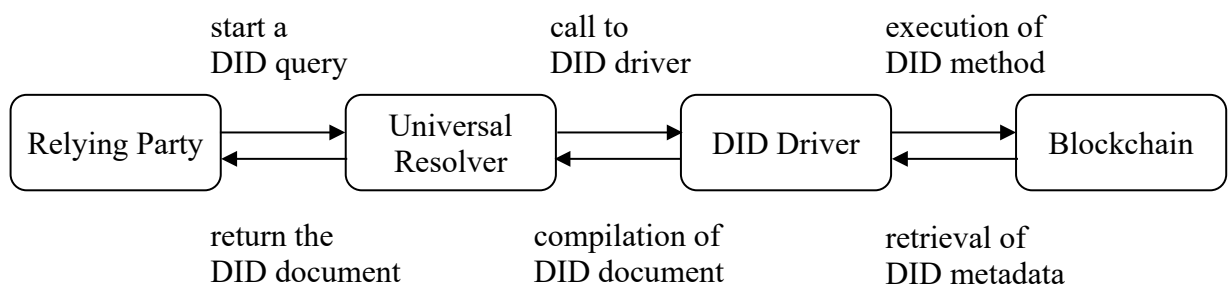

**Figure 6: DID Lookup Using the Universal Resolver**

**Identity Hubs – DIF**:

Identity Hubs [18] are offchain encrypted personal datastores connected together and linked to a given entity. They can be used to securely store identity data (directly on user devices or on user-designated cloud storage providers) and granularly share it when such sharing is approved by the owner. Each Identity Hub is linked to a given DID and can be integrated with the Universal Resolver. A key property is that the data attached to the DID of an entity is replicated and stored across the set of Identity Hub instances run by the entity.

**DID Auth – RWOT**:

DID Auth [19] is a conceptual authentication framework that allows a subject to prove ownership of a DID through a challenge-response cycle that includes the resolution of a DID to its DID document. Multiple architectures exist to support DID Auth interaction cycles; they rely on exchanges of JSON Web Token (JWT) and/or JSON-LD objects that are transported using different combinations of protocols and mechanisms, such as hyper-text transfer protocol (HTTP), near-field communication (NFC), and quick response (QR) code scanning.

**Open Badges – Mozilla and IMS Global**:

Open Badges [20] is another approach to credentials, which are referred to as "badges." There are three core data classes used to instantiate a badge:

- The "Assertion" class contains information about the subject, the issuance timestamp, and instructions to verify the provided assertions. Additional properties can also be added, such as a revocation status or an expiration date.
- The "BadgeClass" class adds context to the credential type (e.g., category), the criteria used to describe how to earn the credential, and a reference for the issuer.
- The "Profile" class adds information about the the badge issuer, recipient, and/or endorser (e.g., name, email address, phone number, public keys).

Badges take the form of JSON-LD documents that can be encoded into QR codes, allowing for easier integration into applications.

## 3.4    Building Blocks

The building blocks of blockchain-based identity management systems vary, but at a high-level, they are commonly comprised of the following technical components.

**Blockchain**:

A blockchain is used to support the mapping of keys to identifiers by acting as an integrity-protected "bulletin board" for *public key infrastructure* (PKI).[1] Blockchains may be application-specific, such as Hyperledger Indy [24], and/or may support native smart contract platforms. In most cases, this "bulletin board" for PKI—sometimes augmented by separate protocols atop the blockchain—forms an identifier management system (called DID method if it follows the DID specification). In addition to keys and identifiers, credentials may also rely on the blockchain.

---

[1] NIST Special Publication (SP) 800-32 [21] defines a PKI as follows: "[A PKI] binds public keys to entities, enables other entities to verify public key bindings, and provides the services needed for ongoing management of keys in a distributed system." Note that the company Evernym was awarded a grant from the U.S. Department of Homeland Security in 2017 to develop a decentralized key management solution based on NIST SP 800-130, *A Framework for Designing Key Management Systems* [22]. This became the key management foundation of the Sovrin [23] IDMS; the Sovrin codebase was then added to the Hyperledger Foundation open-source projects under the name of Hyperledger Indy.

**Second Layer Protocol**:

An identifier management system may rely on both a blockchain and a separate protocol on top of it, often referred to as a *second layer protocol*. These protocols can be used to build scaling solutions by offloading operations away from the blockchain layer. That way, smart contracts can be designed such that blockchain transactions (triggered by function calls) track a set of operations rather than a single one. For example, the SideTree protocol [25] (run by SideTree nodes that are separate from those of the underlying blockchain) allows one to bundle DID operations together before posting them onto a blockchain.[2]

In addition to the scalability benefits, second layer protocols can help promote the development of interoperable, blockchain-agnostic systems. They do not function as standalone blockchains but rather require one or more blockchains to operate, allowing for the integration of multiple blockchains without necessarily involving any fundamental change to their codebase. Finally, second layer protocols may have a different level of privacy than transactions in the underlying blockchain, though this has to be determined on a case-by-case basis.

**Smart Contracts**:

Blockchains may support smart contracts, which are vital to many blockchain-based IDMSs (some implement all logic in the form of smart contracts). The power of smart contracts is that they can implement data processing logic while the blockchain network guarantees its execution. This enables blockchain-based IDMSs to use smart contracts to support novel governance models for credential service providers. In particular, they are currently used to implement onchain registries with various permissioning structures.

**Credential Storage Methods**:

A foundational architectural feature for blockchain-based IDMSs is the method (or methods) by which credentials are stored (see Section 4.4.2 on *Credential Architectures*). Some blockchain-based IDMSs allow for storage of credentials using a blockchain while others store the credentials offchain. Offchain credentials may be stored by a subject in a wallet application (explained in the *User-Controlled Identity Wallet* paragraph below) or by a third-party custodian to whom the subject has delegated this role.

---

[2] The SideTree protocol (released as a DIF project) has been implemented to develop DID methods by Microsoft on top of the Bitcoin protocol (the DID method is called Identity Overlay Networks [26]) and by Transmute Industries (with ConsenSys) on top of the Ethereum protocol (the DID method is called Element [41]).

**User-Controlled Identity Wallet**:

A user-controlled identity wallet is an application that primarily allows a user to hold identifiers and credentials by storing the corresponding private keys. It also serves as an interface for entities to interact with one another. For example, the subjects can receive and approve credentials from the issuers and disclose presentations to relying parties. Actions can be initiated automatically through application programming interfaces (APIs) that may be triggered by a user through scanning QR codes. In certain systems, a user can fully originate identifiers offline, on their own, and directly in their identity wallet (see Section 4.4.1 on *Identifier Architectures*).

Identity wallets may take various forms, such as dedicated hardware wallets, mobile applications, or even paper wallets (private keys are simply printed out and kept in a safe location). They may also come natively in a browser, an operating system, or as extensions. Wallets that are proposed "as a service" by a third-party holder that controls a user's private keys are called *custodial wallets*.

In addition, identity wallets can act as control centers since entities can receive and decide whether to approve requests for verifiable information, thereby giving their consent to perform some action. They may also serve as gateways for accessing and using applications and services. See Section 6.4 on *Ecosystem Convergence* for more insight about interoperability and cryptographic schemes used to secure custodial wallets.

**User Profile Data Management Protocols**:

After being authenticated, users typically create data when using applications (e.g., user settings, browsing data, transaction history). Traditionally, applications maintain user accounts to store and manage that data, and blockchain-based authentication does not necessarily change that. An alternative to that model, however, may potentially be found in the use of external protocols for managing user profile data with blockchain-based IDMSs to control access rights to that data. With such protocols, user profile data created when using an application would not have to be stored by the application itself. Instead, users could rely on encrypted vaults and decentralized storage protocols [27] and bring the data to the application the next time they use it. For example, 3Box [28] provides a protocol for Ethereum-based user profiles that uses the Inter-Planetary File System (IPFS) [40] for data storage and OrbitDB [79] for data management.

**Data Exchange Models**:

To request, issue, disclose, and verify credentials and/or presentations (e.g., for authentication), blockchain-based IDMSs commonly leverage data exchange formats such as JSON, JWT, Security Assertion Markup Language (SAML), and eXtensible Data Interchange (XDI).

**Application Libraries and Interfaces**:

There exist application libraries and APIs that facilitate the integration of applications supporting various identity management roles (e.g., requester, issuer, relying party, and verifier roles). For example, Hyperledger Aries [29] is a framework released by the Hyperledger Foundation that offers several client-side components and wallet services integration to support interactions between participants in blockchain-based IDMSs.

## 3.5   Blockchain Identity Management Stack

The Decentralized Identity Foundation published the draft protocol stack [30] shown in Table 1— another breakdown of blockchain-based identity management building blocks in the form of a layered stack to facilitate the emergence of portable and interoperable solutions. Adjacent layers do not have to be built as separate applications and can be grouped together if desired for simplicity, scalability, or to more closely align with adopted standards. While DID-specific, the stack might be similar for approaches using other identifier management systems.

**Table 1: Proposed Identity Stack (from the Decentralized Identity Foundation [30])**

| Layer | Description |
|---|---|
| Application | Applications that interact with a given identity management system through library integrations and API calls |
| Implementation | Libraries that integrate the system in third-party applications |
| Payload | Message formats, such as JWT, used to exchange data between participants |
| Encoding | Methods for encoding data at both the encryption and payload layers |
| Encryption | Methods for encrypting messages between participants as well as encrypting the data held by the identifier owner |
| DID Authentication | Methods to authenticate a participant using their DID |
| Transport | Transport protocols used for sharing data between participants and devices, such as HTTP or QR code |
| DID Resolution | DID Resolver used to convert a DID into its corresponding DID document |
| DID Operation | Create, Read, Update, and Delete operations for a DID document |
| DID Storage | Methods for storing DID Documents and DIDs |
| DID Anchor | Networks that serve as medium for DIDs |

## 4    Blockchain Identity Management System Taxonomy

This section discusses how blockchain identity management systems are constructed and what differentiates the various approaches. Section 4.1 examines system authority models, identifier origination schemes, and credential issuance schemes. Section 4.2 evaluates methods for identifier and credential management, and Section 4.3 considers presentation disclosure. Section 4.4 looks at different system architecture designs, and Section 4.5 discusses the use of public registries and related implications. Section 4.6 concludes the taxonomic analysis with a higher level discussion of system governance options.

### 4.1    Authority Model

This section discusses the different control models for blockchain-based IDMSs and the different ways for such systems to establish new identifiers for their users.

### 4.1.1    Top-Down vs Bottom-Up Organizational Structures

The authority model of a system specifies how it is controlled. The two main approaches are top-down and bottom-up, with the latter frequently associated with *self-sovereign identity* (SSI) principles [31]. These two authority model approaches form a spectrum, which can support different types of governance structures and power delegation mechanisms.

**Top-Down Approach**:

A system owner acts as a central authority that has control over identifier origination and/or credential issuance. Power may be delegated through roles to create a hierarchical structure. This model may be appropriate for organizations that want to explore distributing their processes and architectures to better meet their needs and provide enhanced control and privacy for the users while keeping ownership of the system and control of its governance, as discussed in Section 4.6. An example of a system that uses this approach is described in *Smart Contract Federated Identity Management without Third Party Authentication Services* [32].

**Bottom-Up Approach**:

No single entity acts as a central authority that has control over identifier origination and/or credential issuance. Participants manage their own identifiers and credentials without requiring any permissions, though they must still follow the rules of the IDMS (often enforced through a set of smart contracts). This does not exclude the possibility of some entities playing more significant roles than others in designing and maintaining the system architecture and incentives.

### 4.1.2   Identifier Origination Schemes

There are many possible methods for *originating* new identifiers within blockchain-based IDMSs. Identifier origination always starts with the generation of blockchain addresses directly by users who control the custody of the associated private keys. In general, this is achieved by generating a public-private key pair and deriving a blockchain address from the public key using a cryptographic hash function and protocol-specific transformations. Blockchain addresses alone, however, do not fully meet the need of identity management; there must be additional onchain logic and registries to use them as identifiers in a given IDMS (that is, subjects may need to make a smart contract aware of their identifier).

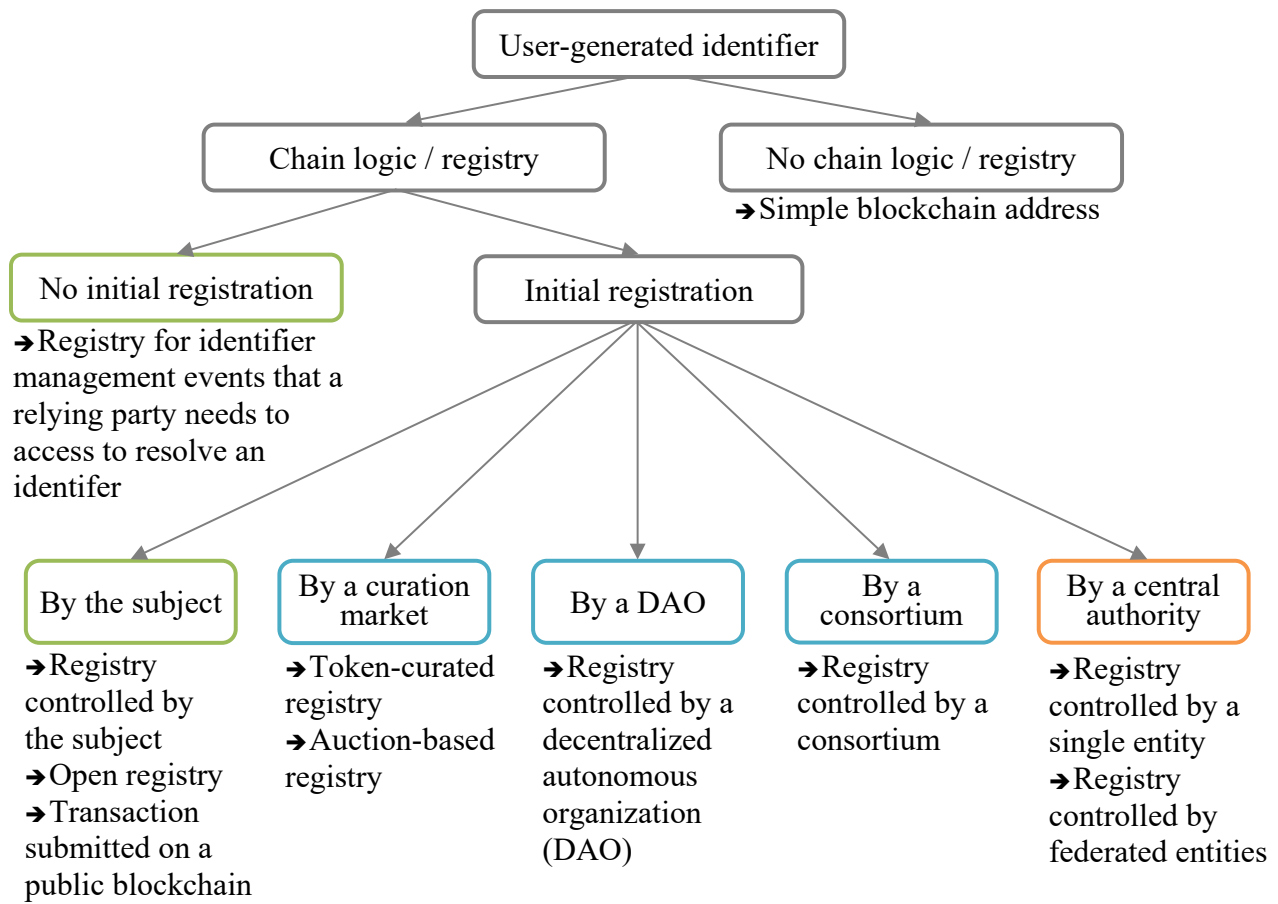

**Figure 7: Identifier Orignation Schemes**

Figure 7 contains a diagram showing different methods that can be used by systems to originate identifiers. Identifier origination based on a central authority with a top-down approach is shown in the bottom right (in orange). Schemes that involve no initial registration (registration only needed at a later date and under certain conditions) or a self-registration following a bottom-up authority approach are on the left (in green). Finally, schemes that involve a curation market (see Section 4.5 on *Public Registries and Reputation Management Implications*), a decentralized autonomous organization (DAO), or a consortium can lean towards one side or the other depending on how the permissions are implemented and controlled by the participants (in the middle of the figure in blue).

An example of DAO-controlled identifier registration for Internet Protocol (IP) addresses can be found in [33]. Section 4.4.1 on *Identifier Architectures* provides different approaches for implementing these identifier origination schemes.

### 4.1.3   Credential Issuance Schemes

A credential is issued to a subject by an issuer following a request by a requester. The approval of the subject may be required, and the issuer may be compensated for issuing the credential (e.g., through some marketplace mechanism built into the protocol).

With the top-down authority model, credential issuance may be controlled or regulated by a central authority (see Section 4.1.1). In the bottom-up authority model, any user can issue a credential to another user.[3] A credential might also be self-issued by a subject; for example, a subject could self-issue a credential to publicly share contact information about themselves (e.g., a public key, a service endpoint) or consent preferences that help other users know how to interact with them.

### 4.2   Identifiers and Credentials Management

This section discusses lifecycle and custody issues related to identifiers and credentials. This includes creation, issuance, discoverability, transferability, recovery, suspension, and revocation.

### 4.2.1   Lifecycle

**Lifecycle Determination at Origination**:

The lifecycle of a given identifier or credential can be set at the time of origination such that there will be no need for outside intervention in the future (e.g., making it expire after a certain amount of time or making it irrevocable). This can enable an identifier or credential to take a lighter, self-supporting form in order to let the subject be more independent (see *Bring-Your-Own Blockchain Address* in Section 4.4.1.2 and *Offchain Object* in Section 4.4.2.2). In the case where the identifier or credential is irrevocable, a relying party may not need to be actively connected to the identity management system in order to verify the credential or identifier. Alternatively, if an identifier or credential does not have its lifecycle fixed, entities need access to the blockchain to verify them.

**Suspension and Revocation**:

An identifier or a credential may be suspended or revoked by the issuer, the holder, or when predefined conditions are met.[4] If an identifier expires, the credentials that are associated with it may not be verifiable anymore. Verifiers are usually expected to verify whether an identifier or credential is expired.

Furthermore, performing these actions may require approval from the participants involved.

---

[3] There are additional advanced schemes to issue credentials anonymously and without relying on any trusted issuer by using the techniques in [34], but the claims for which these credentials are issued must be verifiable by anyone participating in that system. Another credential issuance scheme involves using a threshold of mutually distrusting parties as in [35].

[4] Blockchains can help make the revocation process more transparent and secure; for instance, CertLedger [36] is a scheme that is comparable to Google's Certificate Transparency (CT) while preventing the "split-world" attack that is possible against CT.

Implementation of suspension and/or revocation mechanisms depends on the scheme through which identifiers are registered and credentials issued. Section 4.4.3, *Offchain Objects Coupled with Global Credentials Registry*, offers an example of architecture that incorporates a public *revocation registry*.

### 4.2.2   Custody and Delegation

This section discusses the custody and delegation processes for identifiers and credentials, including ownership, storage, and transferability. Identifiers may be registered onchain or may remain privately stored and used offchain, depending on the IDMS (see Section 4.4.1 on *Identifier Architectures*). In both cases, users can self-custody the private keys and data associated with their own identifiers and credentials. That said, they can also choose to delegate control over them to a custodian for a certain period of time or indefinitely.

Users who lose their private keys may recover them through mechanisms that must be put in place preemptively: a custodian designated by the user, a list of user-appointed trustees (social recovery), time delay mechanisms, and/or a central authority. If there is no such mechanism in place prior to a loss, the private key cannot be recovered, and it is computationally infeasible to regenerate it. Custodians can form marketplaces to provide services while acting on behalf of the subject, such as storage, management of control and consent preferences and relationships with relying parties, recovery mechanisms in case of loss, and authenticated communication channels. Also, an identifier may be abandoned and what is owned by the identifier transferred to another. This may be done for key rotation purposes and not just when the private keys are lost. An identifier that is abandoned without having previously transferred the credentials that are associated with it to another identifier may cause the loss of those credentials.

Some types of credentials may be transferable from one subject to another; this can fit use cases such as representations of ownership (e.g., a certificate proving ownership of a good that an entity may then be able to transfer on their own if and when selling the good), as shown in *Exchanging Concert Tickets and Coupons* in Section 7. Systems can implement this using some form of non-fungible token (see Section 4.4.2 on *Credential Architectures*).

Figure 8 provides a diagram of different interactions between entities and an identity management system; these interactions are either direct or delegated through custodians.

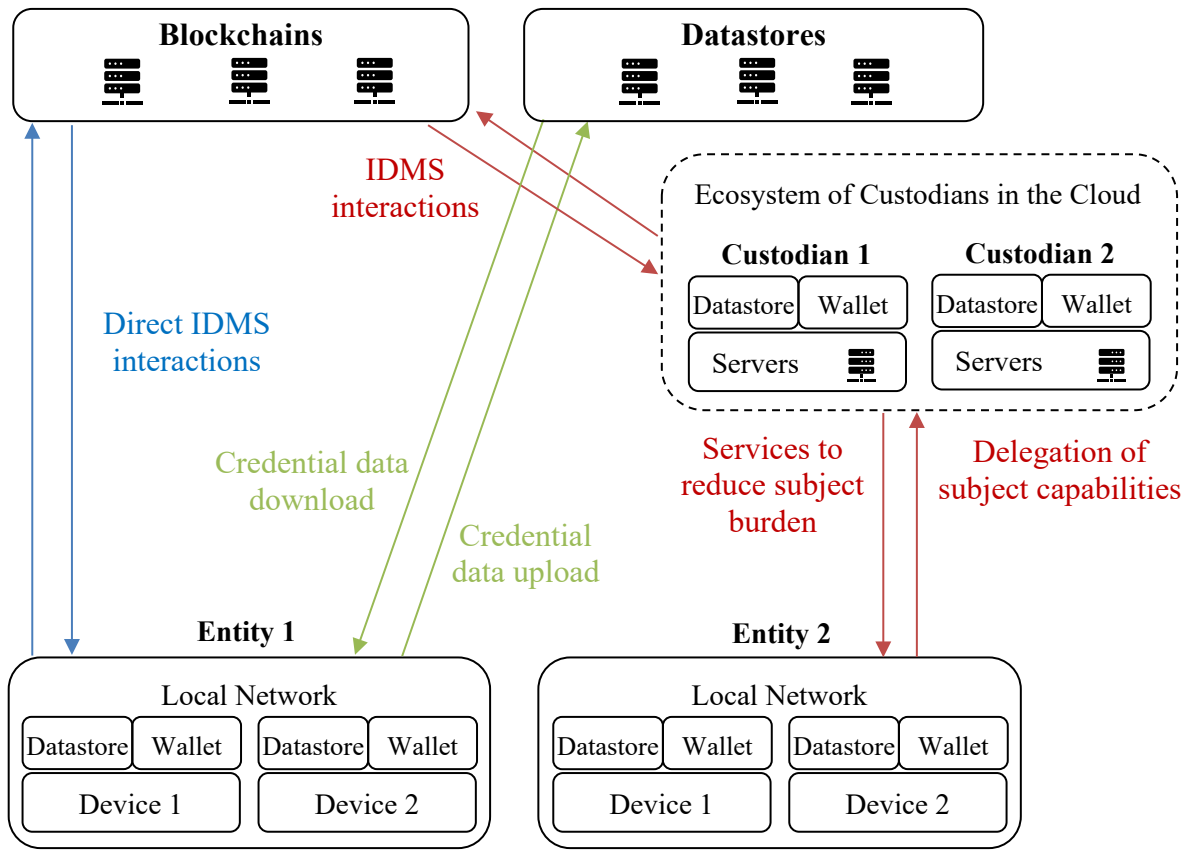

**Figure 8: Interactions Between Subjects, Custodians, and Blockchains**

## 4.3   Presentation Disclosure

A presentation derived from one or more credentials (see terminology in Section 3.1) allows subjects to share verifiable information directly with a relying party and authenticate themselves. The sharing of a presentation from a subject to a relying party is called *presentation disclosure*. This relationship comes with its own management, control, and consent considerations, which the following properties attempt to characterize.

Subjects can control the release of their data with relying parties (e.g., businesses, applications) and may do so at differing levels of granularity to limit information being released to the minimum necessary. Subjects may be compensated for presentation disclosures, thereby monetizing their own data.

**Selective Disclosure Mechanisms**:

A presentation disclosure may involve sharing an entire credential, one or more claims from a credential, or a quality derived from a credential. A presentation can include a minimal amount of information to interact with a relying party on a need-to-know basis with a *zero-knowledge proof* for verification, potentially providing subjects with the ability to avoid oversharing personal data.

Zero-knowledge proofs are cryptographic schemes in which a prover is able to convince a verifier that a statement is true without providing any more information than that single bit (that is, that the statement is true rather than false).

Consider a patron who is stopped by the bouncer while attempting to enter a bar because the bouncer must be convinced that the patron is at least 21 years old. The patron shows the bouncer their driver's license, the bouncer quickly looks for a birthday, and then the patron can enter if they are of age. In this scenario, the bouncer learns far more information about the patron than would be ideal, and a particularly malicious bouncer may be able to learn enough about the patron that they can commit identity theft. Contrast this example with one that employs a zero-knowledge proof scheme. The prover (the patron) proves to the verifier (the bouncer) the statement "I, the prover, am at least 21 years old." They are able to do so without revealing their birthday, driver's license number, or any other information. The patron then enters the bar with their identity and privacy secure, but a different, underage patron is unable to create a convincing proof.

Zero-knowledge protocols (those utilizing zero-knowledge proofs) are a major area of active research (see Section 6.2 for a high-level technical overview).

Note that there exists selective disclosure mechanisms that do not involve zero-knowledge proofs. For instance, Merkle trees can be used to structure credential data, allowing subjects to disclose select parts of the tree to verifiers who can then check the validity of the data with a Merkle proof (e.g., as found in Bloom [60]).

**Pairwise-Pseudonymous and Single-Use Identifiers**:

Users may be able to maintain a set of special purpose identifiers that are not linked to the primary identifier, which enables users to maintain a level of anonymity. For example, participants may use pairwise-pseudonymous identifiers where they have a unique, dedicated identifier for each relationship they have with a third-party. Alternately, they may use single-use identifiers that are discarded after a particular interchange [37].

*Hierarchical deterministic* (HD) wallets, specified under the Bitcoin Improvement Protocol (BIP-32 [38]), can be used to derive unlinkable identifiers (referred to as *children accounts*) from a single master key. Note that identifier unlinkability schemes can be combined with selective disclosure mechanisms.[5]

**Unicast, Multicast, and Broadcast Disclosure Modes**:

A presentation can be disclosed to a single relying party, a group of relying parties, or everyone. Public disclosure has reputation management implications (see Section 4.5) and is often used by relying parties who publicly disclose a presentation about themselves in order to prove who they are and justify that they have a valid reason to request presentations and receive personal information from participants.

---

[5] For example, [39] presents a system built atop Bitcoin that uses Brands' commitment scheme to let users selectively disclose their credentials via zero-knowledge proofs.

**Efficiency and Cost**:

A presentation may require onchain processing at the time of disclosure by the subject. Alternatively, a self-contained presentation can be disclosed by the subject without interacting with the blockchain. The relying party receiving the self-contained presentation may still need blockchain access to process and verify it. These considerations result in solutions with varying efficiency and costs. Some actions can be achieved offchain, quickly, and with no transaction fee. Others may be free of cost but require access to the published blocks. Finally, actions may be delayed by transaction processing time and induce costs for paying the blockchain miners to process a transaction.

In the case of smart contract-based systems on top of permissionless blockchains, third-party entities may pay smart contract transaction fees on behalf of the users so that they do not have to deal with holding and spending the native digital currency of the blockchain themselves.

## 4.4 System Architecture Designs

This section focuses on the architectural designs that can be followed when building a blockchain identity management system. Pieces of the system can be constructed as distinct modules or combined into monolithic architectures, though some designs are mutually exclusive. Some of them use onchain registries and logic (generally implemented in the form of smart contracts on a blockchain that can be, depending on the purpose, permissionless or permissioned) that may be augmented by system-specific offchain schemes.

Architectures for identifiers are first discussed, then credentials, and finally, more complex combinations of architectures for identifiers, credentials, or both.

### 4.4.1 Identifier Architectures

This section discusses the technical means to implement the identifier origination schemes introduced in Section 4.1.2.

#### 4.4.1.1 Onchain Registry

**Credentials Registry Acting as Identifier**:

For each identifier participating in the system, a dedicated smart contract is deployed that can store credentials for that identifier; this architecture typically follows a bottom-up authority model approach.

The deployment of a new contract for every identifier allows participants to have control over their own identifiers by enabling them to prove ownership of that contract. This can come at higher costs since many contracts must be deployed, more data must be posted on a blockchain, and there may be slower processing speeds due to the number of transactions on a blockchain. Interoperability issues can arise if different identity management contracts are deployed by different users (or simply different versions of the same contract). This may be mitigated by using standards such as ERC-725, *Proxy Account,* a proposed Ethereum standard that follows this architecture. It allows other smart contracts to take action based on verifiable identity information contained in ERC-725 smart contracts. In addition, ERC-734, *Key Manager*, can complement them by allowing subjects to delegate certain capabilities to custodians of their choice.

**Global Identifiers Registry**:

A single monolithic smart contract or set of integrated contracts is deployed that acts as a global registry for storing and managing all identifiers; this approach can follow either the top-down or bottom-up authority models. The writer of the smart contract can encode a variety of possible governance models wherein the entity deploying the contract has complete control of the system, limited control of it, or no control of it. In the case of no control, the governance of the contract would have to be run by participating users (e.g., with a DAO). The registry can contain all the necessary logic and data to resolve identifiers to their metadata (e.g., DID documents when the DID specification is followed) or contain only hashes that are mapped to the actual metadata stored elsewhere.

**Anchors Registry**:

A single monolithic smart contract is deployed that acts as a global registry that registers the hashes of identifier management operations that are grouped together into bundles or *anchors*. The bundling (i.e., grouping) of identifier management operations is executed by a second layer protocol that sits on top of the blockchain in which the anchors registry is located. The protocol then adds the hashes of those anchors in the registry and uses decentralized storage systems, such as IPFS [40], to store the anchors data (identifier management operations). The Element [41] identifier management system, based on the SideTree protocol on top of the Ethereum blockchain, follows this architecture.

Note that an anchors registry (coupled with a second layer protocol) may be used for any onchain registry (e.g., to support credentials instead of just identifiers).

### 4.4.1.2    Bring-Your-Own Blockchain Address

Any blockchain address is a valid identifier and can be immediately used without having to be registered beforehand. Identifier creation and storage is usually done locally in the identity wallet. This architecture follows a bottom-up authority model where the user is self-reliant. Identifier creation takes place offline without gatekeepers or transaction fees.

Identifier management (by the subjects) and authentication (by the verifiers), however, may require onchain capabilities. This differs from the identifiers' registry smart contract architecture because identifiers are not initially registered and stored onchain, making them non-discoverable by default.

This architecture may help the system operate at scale since no blockchain transactions are needed for initial identifier creation. Users control their identifiers, as with the per-identifier smart contract architecture, and may gain privacy advantages since identifiers need not be publicly viewable. Moreover, lightweight identifier creation could facilitate the use of pairwise pseudonymous identifiers (or unique, one-time identifiers) to enhance their privacy when interacting with relying parties (see *Pairwise-Pseudonymous and Single-Use Identifiers* in Section 4.3). Onchain logic may be necessary to implement additional functionalities such as identifier management capabilities.

For example, it is needed to access the chain to resolve an identifier. The information necessary to do this must be stored on a blockchain and likely managed through some smart contract.

ERC-1056, *Lightweight Identity*, is a proposed Ethereum standard that follows this architecture and is used by uPort's DID method called "ethr" [42]. DID operations are stored in the form of Ethereum events. Resolving a DID to its DID document consists of iterating over the DID operations that may have been posted by the subject. Protocols that define and implement DID methods to build DID documents for bring-your-own blockchain identifiers may be further developed in a way to interact with multiple blockchains. Note that, in the case of Ethereum, blockchain log data cannot be queried from other smart contracts; however, an external method can be designed to access the chain and iterate over the logs to build a document (as in uPort).

### 4.4.1.3   Unspent Transaction Output Model

Architectures can also rely on the *unspent transaction output model* (UTXO) that certain blockchain protocols follow (e.g., Bitcoin). Identifiers are created by submitting blockchain transactions using newly generated blockchain addresses. The unspent transaction outputs of those transactions are then used to inform and manage the identifiers' statuses.

The Bitcoin Reference (BTCR) DID method [78] is a proposed DID method that follows this architecture on top of the public Bitcoin blockchain. The information contained in the unspent transaction output of BTCR transactions is used to resolve DIDs and build DID documents.

### 4.4.2   Credential Architectures

This section discusses architectural designs for storing and managing credentials. The choice of design may depend on how identifiers are managed. There are both onchain and offchain credentials storage and management methods, though they present different usability, privacy, and security implications.

Onchain credentials often only require onchain storage for the hashes of the credentials with the non-hash data being stored on any data store a subject has access to, be it a designated custodian or a decentralized storage system such as IPFS [40].[6] The integrity of the data may be checked by the receiving party by hashing the credential and comparing the hash with the one found on the blockchain. The hashes are often stored as state variables or blockchain logs, the latter being sometimes cheaper than onchain storage (e.g., Ethereum events).

Credentials can also be stored fully offchain, either directly on the subject's device and/or by a designated custodian. There may still be, however, onchain mechanisms to handle revocations and other credential status updates.

---

[6] In addition to using IPFS with an onchain pointer, the research literature has demonstrated a number of designs for how to store credentials (and other data) offchain securely. For example, [43] uses a blockchain for enforcing access control policies on an offchain data store, which is implemented as a distributed hash table (such as Kademlia). An alternative system described in [44] uses centralized and decentralized databases linked together by a blockchain for data write access management while providing external queryability (which can be useful for areas such as medical research). Finally, Calypso [45] is a more advanced construction with auditable access control, which uses threshold cryptography to protect access to data.

#### 4.4.2.1    Onchain Registry

**Per-Identifier Credentials Registry**:

In this architecture, credentials are managed as entries in a per-identifier smart contract that acts as a container as defined in Section 4.4.1.1. This architecture can give the subject unilateral control over their credentials. As owner of the contract, a subject can remove any credential they want without the approval of the credential issuer. Also, their approval is required, in addition to that of the credential issuers, for the issuance of a credential (see Section 4.1.3).

While subjects can manage their own onchain credentials in this way, this architecture is heavily reliant on onchain transactions. This can hinder system scalability due to blockchain transaction costs and the relatively slow processing speed for transactions. The architecture can thus make it expensive to use privacy features, such as pairwise pseudonymous identifiers, for every relationship (see *Pairwise-Pseudonymous and Single-Use Identifiers* in Section 4.3). ERC-735, *Claim Holder*, is a proposed Ethereum standard that follows this architecture and can be utilized jointly with ERC-725, *Proxy Account*.

**Global Credentials Registry**:

In this architecture, credentials are registered and managed as entries in a single smart contract. Similar to the *global identifiers registry* architecture, the identifier that deployed the contract initially owns the system. However, that authority can be delegated, transferred, or limited depending on how the contract is coded. Thus, this architecture requires the initial owner to set up a governance model that establishes the rules and permissions for managing credentials. This may necessitate handling concepts such as reputation and negative credentials (see Section 4.5).

Another key aspect is that credential management involves onchain transactions and access, which impact the usability and cost of presentation disclosure (as discussed in Section 4.3) as well as privacy (as discussed in Section 5).

This architecture can be used as a registry for revoking credentials. A relying party is then able to verify the validity of the offchain credential. Another use for this architecture is to allow a user to publish credentials about themselves and share information publicly, such as a public key or a service endpoint.[7] ERC-780, *Ethereum Claims Registry*, is a proposed Ethereum standard that follows this architecture. ERC-1056 (see Section 4.4.1.2 on *Bring-Your-Own Blockchain Address*) also implements a credentials registry, though it is limited to self-issued credentials and is based on blockchain logs.

**Non-Fungible Token Registry**:

In this architecture, a credential takes the form of a *non-fungible token* (NFT). An NFT is a unique, non-interchangeable token that is owned and may be transferable. Minting and management of the tokens are performed through an NFT factory smart contract that acts as a registry that manages the NFTs. NFT-based credentials primarily fit use cases that deal with digital ownership, especially—but not exclusively—when it is meant to be transferable (see Section 4.2.2 on *Custody and Delegation*).

---

[7] Advanced cryptographic primitives, such as the hash-based accumulator employed in [46], can allow a registry to retain a constant-sized storage regardless of how many credentials are registered.

The minting of specific tokens can implement application-specific token formats, rules, and requirements and therefore provide dedicated token lifecycle management capabilities. In addition, this architecture can use interoperable token formats, thus enabling a marketplace for transferable credentials. NFTs can either be issued individually or to a group (a distribution method also called *airdrop*). These capabilities come at the expense of the need for participants to issue blockchain transactions and have blockchain access (See *Efficiency and Cost* in Section 4.3).

ERC-721, *Non-fungible Token Standard*, is a proposed Ethereum standard that follows this architecture. As an example, 0xcert provides a framework for building applications that create and manage ERC-721-compliant NFT-based credentials [47].

**User-Mintable, Predefined, Non-Fungible Token**:

In this architecture, a credential takes the form of an entitlement to let a user mint a predefined and pre-assigned NFT at a future date or condition. This can be achieved through system-specific NFT factory smart contract designs. As an example, Centrifuge [48] allows one to turn offchain credentials into NFTs by building proofs that are verified through Merkle root hashes (stored onchain) of some of the offchain credential data.

This may also be achieved for a group of subjects through the use of a *Merkle airdrop* (see definition in Glossary in Appendix B), which allows group distribution of the entitlement to redeem an NFT. This scheme is highly scalable in that it requires only one transaction by the issuer and is independent of the size of the group. No management support is needed after the distribution as all of the activity comes from the subject side.

A credential is private by default, and a subject can redeem it only if they want to use or transfer it. However, the list of all the identifiers the Merkle airdrop was issued to must be available to the subjects to redeem their NFT (both the private key and the list of all the identifiers included in the Merkle airdrop are needed to build the Merkle proof and mint the NFT). In Merkle airdrops, tokens must be *pulled* by the users while for traditional airdrops tokens are *pushed* to them without any approval needed.

#### 4.4.2.2    Offchain Object

In this architecture, a credential takes the form of an offchain object that acts as a self-contained vehicle for transmitting information directly between parties.

This can go hand in hand with the *bring-your-own blockchain address* architecture (discussed in Section 4.4.1.2) to establish a lightweight identity management system that can operate at scale. It best matches use cases where the lifecycle of a credential is predetermined. However, verification of a credential (see *Lifecycle Determination at Origination* in Section 4.2.1) may require chain access (see *Efficiency and Cost* in Section 4.3). In particular, if revocability is permitted, onchain artifacts are required for one to check the credential revocation status, such as credential revocation registries (see *Offchain Objects Coupled with Global Credentials Registry* in Section 4.4.3).

It can provide a high level of control and autonomy to the subjects as they can manage the storage of their own credentials offchain. It ensures privacy by default and need not be constrained to a specific blockchain. This architecture may use, for example, the JWT format (see Section 3.4 on *Building Blocks*), as in Blockstack [49].

### 4.4.3    Combination Patterns

It is possible to combine the architectures for identifiers, credentials, or both. This section provides some examples of how this can be done but is not exhaustive.

**Global Identifiers Registry Coupled with Per-Identifier Credentials Registry**:

An IDMS can be designed so that identifiers are stored in a global registry, though each identifier has its own dedicated smart contract for storing and managing credentials. The Smart ID project from Deloitte [50] follows this architecture. The global identifiers registry may also serve as a smart contract factory to create and manage all of the per-identifier credentials registry smart contracts.

**Global Registry for Both Identifiers and Credentials**:

A single smart contract can implement both an identifiers registry and a credentials registry as described in Sections 4.4.1.1 and 4.4.2.1.

This approach is followed in *Smart Contract Federated Identity Management without Third Party Authentication Services* [32]. Another example of this approach is smart contract-based PKI (SCPKI) [51], which stores all identifiers and credentials on a single smart contract and allows relying parties to use a web-of-trust to decide whether or not an identifier is authorized to perform some action. SCPKI can also be extended with blind signatures in order to provide privacy [52].

Another example is BlockPKI [53], which can generate one or more smart contracts per identifier in the system. These per-identifier contracts (called "certificates") contain a set of credentials and are used to store signatures from certificate authorities; once enough signatures have been gathered in a contract, they are aggregated and sent with the certificate data to a global credentials registry contract. Relying parties can use this global credentials registry to verify signed certificates in the system.

**Offchain Objects Coupled with Global Credentials Registry**:

Offchain objects can be used as the primary way to issue and share credentials while relying on a central registry smart contract to publicly store information necessary to exchange those credentials (e.g., service endpoint URLs, public keys).

A credentials registry can also be leveraged to act as a revocation registry for offchain credentials. Such a registry is used in both uPort [42] (it is based on ERC-780 and deployed on the public Ethereum blockchain) and Hyperledger Indy [24]. In the latter, an issuer can control a revocation registry that relies on a *cryptographic accumulator* (i.e., a protocol that allows one to prove a membership in a set; see Section 6.2) to allow relying parties to verify whether a given credential was revoked by the issuer without compromising the registry's security and the holder's privacy.

**Offchain Objects Coupled with Global Identifiers Registry for Issuers**:

Issuers have their identifiers stored on an onchain registry. They can issue offchain credentials directly to any blockchain addresses controlled by the subjects. Verifiers only need to verify that the signatures of the credentials issuers match those on the onchain registry.

**Non-Fungible Tokens with Global Credentials Registry**:

Rules and permissions based on a global registry smart contract can be implemented to restrict the context in which transfers of NFT-based credentials take place. This way, parties that trust each other can transact securely and according to the agreed-upon rules.

This can be leveraged, for example, to establish know-your-customer (KYC) checks for exchanges of tokens as in the Transaction Permission Layer Protocol [54] with the ERC-1616, *Attribute Registry*, Ethereum standard proposal. The global credentials registry acts as attribute-based access control and may be used to define granular requirements, such as multi-tiered identity verification associated with different levels of transfer amount and transfer frequency allowances for exchanges of a certain token.

## 4.5    Public Registries and Reputation Management Implications

Some blockchain IDMS architectures rely on onchain registries (e.g., smart contracts) and may contain publicly readable personally identifiable data. This can be leveraged by subjects wanting to share public information about themselves (e.g., a service endpoint at which they can be reached if they wish to be discoverable). It can also be used by organizations wanting to build reputation systems such as public institutions (e.g., TheOrgBook project of the Government of British Columbia [55] running on VON [10], "a public repository of verifiable claims about organizations") and e-commerce platforms (e.g., product and seller ratings).

Publicly readable data does not necessarily imply that user privacy is violated or that users do not have control over their identifiers and credentials. Schemes may use hashing or encryption to protect publicly posted data, and varying degrees of granularity can be implemented to enable users to manage their own credentials and associated reputations.

One important design feature is whether or not user consent is required prior to a credential being issued to that user; the user may view certain claims about themselves as being negative and not want them published. Some systems allow unilateral claim issuance while others require user approval. If the user cannot stop the claim from being issued, they may then want to get a counterclaim issued. A reputation system may be used to track the reputation of issuers, which verifiers can then evaluate. Such systems must protect themselves from attacks designed to inappropriately alter user reputation.

**Sybil Attacks and Structural Barriers**:

Reputation systems need to protect against *Sybil attacks*, in which an attacker pretends to be many people at once, by imposing a structural barrier.

For systems with access control (that may sit on top of either a permissionless or permissioned blockchain), it can take the form of identifier verification and the use of roles and permissions.

For open systems, the structural barrier can be a cost to register, exist in, and/or exit the system. This makes attacking the system disproportionately expensive compared to the benefits the attack would produce. While transaction fees act as a basic cost structure, more advanced barriers that rely on game theoretic concepts can be designed to achieve objectives such as disincentivizing participants from leaving an identity to regain newcomer status and ensuring that participants do not get an advantage by issuing multiple identities.[8]

An example of such a cost structure includes *token-curated registries*, which feature an incentivized voting game to let a community of participants decide whether an entry should be added or removed from the registry. These Sybil-resistance mechanisms can be based on staking funds (e.g., with collateral and/or escrow contracts), reputation, or work (committing a certain amount of resources for a certain period of time).

## 4.6 System Governance

Blockchain-based IDMSs must have a governance structure that makes the system trustworthy to its participants. Approaches can vary significantly and often involve a combination of both onchain and offchain organizational structures.

The onchain structures may consist of smart contracts deployed on some underlying blockchain (either permissioned or permissionless); users are thus required to trust both the governance models of the smart contract-based system and the underlying blockchain. Alternatively, solutions exist where a blockchain is developed and deployed for the sole purpose of supporting an IDMS, called *application-specific* blockchains.

There may be security tradeoffs between these approaches. If the blockchain is not application-specific, governance of the blockchain itself is an important topic but not the focus of this paper, which examines the identity application specifically (a few applicable considerations are provided for reference in Section 6.1 on *Underlying Blockchain Implications*).

---

[8] Reference [56] describes three other types of generic attacks against a reputation system—bad-mouthing, ballot-stuffing, and whitewashing—and proposes a blockchain-based solution to mitigate them. Reference [57] is another blockchain-based reputation system designed for reputation in file-sharing networks or for e-commerce, while [58] aggregates social media reputation.

A set of the higher-level recurring governance traits are discussed below.

**Ownership and Funding**:

A system can be owned by a for-profit organization (e.g., a company), non-profit organization (e.g., a foundation), consortium, government agency, open-source community, and/or DAO. It can be directly financed through traditional fundraising and monetized by the entities that administer it. It can also rely on crowdfunding through an *initial coin offering* (ICO), for example. Tokenholders are not necessarily shareholders of the system in that the tokens may not give any piece of ownership of the system. Finally, the system may have no dedicated funding at all and be maintained solely on a voluntary basis by members of the community.

**Operating Model**:

An IDMS can be designed and administered as a permissioned system to meet the needs of the members of an organization or a group of organizations. This means that only an approved set of users may access and maintain the system.[9] This permissioned system might be offered as a proprietary service to customers or be deployed internally. Access control takes place either at the smart contract level (that sits on top of an underlying blockchain) and/or at the blockchain protocol level (i.e., a permissioned blockchain).

Permissioned blockchains require identity management systems to determine who the validators are (e.g., permissioned blockchains based on a proof-of-authority consensus model). This may take place offchain (typically the validator nodes have a list of the other nodes that they want to connect with) or via smart contracts onchain. Changes to the list of validators may then be administered through onchain voting by administrators.

Alternatively, an IDMS may form an open protocol and/or ecosystem that can be used and integrated by anyone. It can be a general purpose ecosystem or a more application-specific one (e.g., credit scoring with Bloom Protocol [60]). Furthermore, an IDMS can involve users authenticating at the application level or at the ecosystem level such as in Blockstack [49]. The latter differs from traditional single sign-on identity management in that identifier origination, credential issuance, and presentation disclosure are not necessarily controlled by a single entity.

In some systems, tokens may be utilized to design an incentive structure and boost certain desired behaviors from the participants (e.g., through earning rewards) to facilitate ecosystem coordination, self-sustainability, and growth (can be based on various game theory techniques). The incentive structure can be extended to built-in monetization schemes to buy and sell services. More specifically, they may be coded directly as part of the functions that implement actions, such as credential issuance and presentation disclosure.

---

[9] A second layer protocol can be used as an access control mechanism for permissioned blockchains. For example, the ChainAnchor scheme [59] offers this while allowing users to transact pseudonymously and maintain transaction unlinkability: users can selectively disclose their transactions if asked to (e.g., for regulatory purposes) without revealing their other transactions. This scheme makes use of the "Enhanced Privacy ID" zero-knowledge protocol.

**Internal Rules Management**:

Every system has rules that dictate how participants interact with a given system. These rules are often implemented and enforced through smart contract code that is visible to all participants. Since the underlying blockchain enforces correct execution of the smart contract, users can trust that these rules are executed correctly. These rules may also specify how changes to the rules themselves are managed (e.g., how the system is upgraded). Allowing such rule changes may prove beneficial—even necessary—for mitigating security issues or adding new features.

However, allowing arbitrary changes can hinder user trust in the system, especially if those changes can be made without user consent. Thus, the upgradability of these systems usually has to be treated carefully so that expectations regarding the immutability of contracts remain valid [61]. It may be important that there exist platforms to communicate and facilitate decision-making among stakeholders of a system (e.g., to raise awareness of the desired benefits and the associated risks of a certain proposal).

The modifications to the smart contracts can be actively governed by the system's users through a voting system (like Bloom's polling mechanism [60]) or through a DAO. Those modifications may be enforced with a time delay to let participants opt out of the system if they are not satisfied with the rule changes. It could also be possible for a system to have multiple versions live simultaneously (for example, both the upgraded and non-upgraded versions), allowing participants to opt into updating to the new version. Finally, time-stamped entries in an onchain public registry (immutable and tamper-resistant) can facilitate accountability by serving as support for posting update proposals using accounts with identifiers that are part of the same IDMS as the one for which the update is being proposed.

**Software Management**:

The management of the software for a system is a vital governance issue as the software implements the rules and maintains the system. Blockchains can provide significant security advantages for identity management systems, but if the user software is vulnerable, corrupted, or malicious, these protections mean little.

The software can be managed by the developers as an open-source project shared publicly using version control platforms, such as Github, or the software can be proprietary and kept private. Development patterns can be leveraged to enable smart contract upgradability (e.g., a registry contract that points to the latest version of the main contract of the system or an interface contract that is inherited by the system and defines a set of key functions and parameters). Periodic third-party audits, automated tests, and reports can also be performed and disclosed to help assess whether the rules are properly enforced.

**External Influences**:

A given blockchain-based IDMS can be subject to external influences (that usually depend on its operating model) such as:

- Regulatory compliance requirements (e.g., the European Union's General Data Protection Regulation) and law enforcement
- Industry alliances (e.g., Ethereum Enterprise Alliance, Hyperledger Foundation, Decentralized Identity Foundation, Trusted IoT Alliance) and standards bodies (e.g., International Organization for Standardization (ISO), Internet Engineering Task Force (IETF)) that publish specifications, formats, protocols, and patterns
- Peer-reviewed research and bug bounty programs
- Social norms and user expectations

A key implication is that they introduce a certain framework of disclosure and transparency, which might directly affect or even require certain protocol designs. This can help participants be aware of, educated about, and supportive of the rules of the platform. Community expectations may play a significant role in holding the administrators of a system accountable, especially if it is possible to opt out at a reasonable cost and migrate to another provider.

## 5     Security and Risk Management

Blockchains can provide security advantages to a variety of applications by making it easier for users to control the custody of their own identifiers and credentials, thus reducing the reliance on individual third parties. A key implication is that this can provide an alternative to the large databases of user information found in traditional and federated systems for which data breaches can compromise substantial amounts of data at once. More generally, blockchains and the related foundational building blocks discussed in this publication can provide enhanced integrity, privacy, and interoperability. However, they do not solve all issues, and careful examination of the security risks and challenges of blockchain usage is needed.

Some of these issues and associated mitigations are discussed below.

**Private Data Leak**:

When a user shares personal data with a relying party, the relying party may share that data outside of the context of the IDMS. This is a significant problem for any identity management system where personal user data is shared. However, this can be mitigated by the use of minimal presentation disclosure mechanisms. For example, zero-knowledge protocols may be utilized to share presentations that contain only the minimal necessary information for a given interaction with a relying party rather than disclosing full credentials.

Separately, architectures that put less data onchain may, in general, be more privacy-preserving, though it depends on the exact architecture being used and how that data is stored (e.g., unencrypted, encrypted, pointers to outside repositories, hashes). Finally, vulnerabilities may be found in the authentication and messaging protocols used by a given system to support peer-to-peer data transmissions.

**Metadata Tracing**:

Pattern analysis techniques may be applied by attackers to onchain metadata and possible interceptions of messages between parties. For example, they may look at the time that transactions or credentials were submitted to the blockchain, which issuers signed them, or the IP addresses from which they were broadcast. This information may be leveraged by attackers to compromise the confidentiality of *personally identifiable information* (PII). This correlation risk can be minimized by decoupling users from a unique persistent identifier through the use of pairwise pseudonymous identifiers (or more advanced identifier unlinkability techniques). Zero-knowledge proofs may also be used to obfuscate the details of blockchain transactions.

**Replay Attacks and Impersonation**:

A rogue relying party can attempt to collect user credentials and presentations with the intent of fooling another relying party into believing that they are that user. This kind of man-in-the-middle attack can be mitigated through relying parties using certain challenge-response protocols and encrypted tunnels such that the users must always prove their identities (that they know the private key for the identifier associated with the transaction).

**Private Key Compromise**:

In most IDMSs, knowledge of a private key for an identifier is equivalent to owning the identifier. Thus, preventing the compromise of private keys is essential. Keys can be compromised due to errors in key generation, storage, or use, or they can be stolen by malicious actors. Human errors can be mitigated through secure and user-friendly tools for key management and secret sharing (i.e., identity wallets); as discussed in [62], a system may be secure only if it is usable. Identifier recovery mechanisms may be implemented to enable a user to regain control of an identifier in case of loss or theft (see Section 4.2.2.). In general, architectures that provide more privacy may reduce the risk of being targeted and having private keys stolen.

**Data Withholding Attacks and Data Availability Issues**:

When users manage their identifiers and credentials themselves, they benefit from a high-level of autonomy and can ensure the availability of their data. An alternative approach is for users to choose to rely on custodians to hold and manage their data for convenience. However, custodians can misbehave, compromising the ability of users to access their identifiers and credentials. Although properly delegated control restrictions can help constrain such a rogue custodian, this does not prevent data withholding attacks. Even a well-behaved custodian can experience temporary service disruptions (or even go out of business), thus making user data unavailable.

Therefore, it may be important for a user to implement data redundancy by storing multiple copies of identifier and credential data in locations that are either directly controlled by the user (such as identity wallets across different personal devices) or delegated to custodians with proper access and control permissions in place. This could involve identity hubs as mentioned in Section 3.3 on *Emerging Standards*. Note that these are issues with traditional IDMSs and that the use of blockchain can be seen as a potential improvement.

When the user has custody over their data, another potential vector of withholding attacks can stem from a subject misleading a verifier by withholding critical data.

**Quantum Computers**:

Blockchain networks depend on cryptography for their security, particularly on public key cryptography. If a sufficiently powerful quantum computer is built in the future, the most widely used public key cryptographic algorithms in blockchain systems will become insecure. This represents a long-term concern for the security of the private keys used in a given blockchain-based IDMS, which control identifiers and credentials. Note that this concern applies to the web in general; it is not just a concern for blockchain technology. Developing, standardizing, and promoting the use of quantum-resistant algorithms may help alleviate the threats borne by quantum computers.

**Smart Contract Flaws**:

The smart contracts implemented to support the blockchain-based IDMS may have security flaws. Indeed, while smart contracts are usually short and concise, there have been flaws discovered in ones that were published, which enabled them to be compromised.

Audits, tests, formal verification, and the use of well-established libraries can help mitigate this risk. Furthermore, data integrity at the smart contract level may be achieved by establishing permissions to prevent unauthorized participants from accessing and modifying user identifiers and credentials.

**System Governance Design Flaws**:

Some blockchain identity management system architectures (e.g., top-down authority models) may incorporate logic that creates single points of failure. For example, they may provide a certain type of participant a high level of privilege that could be improperly used.

This can be mitigated by instituting appropriate separation of authorities between participants along with a security analysis of the system to identify single points of failure with respect to bad actors in the system. Furthermore, governance architectures that rely on game theoretic incentives have their own risks (e.g., see Section 4.5 on *Public Registries and Reputation Management Implications*).

**Oracles and Second Layer Protocol Compromise**:

A blockchain-based IDMS may integrate offchain data, logic, and processing in the form of oracles and second layer protocols. Should they be compromised, the onchain part of the system may not be able to adequately identify the threat and cope with the compromised data, resulting in a "garbage in, garbage out" situation. It is therefore important to be aware of that possibility and ensure that the necessary checks and balances are in place.

## 6    Additional Considerations

This section provides additional considerations regarding some of the fundamental topics of blockchain identity management previously discussed.

### 6.1    Underlying Blockchain Implications

Blockchains have unique properties and underlying governance implications that must be considered while designing an identity management system or deciding on one to use. Section 7.2 on *Users Involved in Blockchain Governance* of NISTIR 8202 [11] states: "the software developers, publishing nodes, and blockchain network users all play a part in the blockchain network governance." Below are some key considerations on that subject.

**Data Persistence and Privacy**:

Any data added to a blockchain will be permanently available. This can have substantial ramifications for privacy in multiple ways:

- If personal information is encrypted and then stored on a blockchain, confidentiality for that data will be lost if the encryption algorithm is broken.
- Over time, as more and more individual metadata is shared with various relying parties and credential issuers, it can be correlated with onchain data in order to link users and their activities (see *Metadata Tracing* in Section 5 on *Security and Risk Management*).

While the effects of metadata tracking in these systems require more study, the permanence of blockchain data affects anyone who uses a blockchain-based IDMS. That said, there are systems being developed and put into production that may allow for the building of finer privacy schemes.

**Consensus Algorithms, Time Delays, and Data Integrity**:

There are a wide variety of distributed consensus algorithms—both permissioned and permissionless—with different properties that may be important to schemes built on top of the ledgers that use them. A consequence of this is that a scheme built on top of blockchain A may have different security, integrity, and usability considerations than an otherwise identical scheme built on blockchain B. The simplest example of this is the expected delay between broadcasting a transaction and having it included in a block. Permissioned consensus algorithms tend to add blocks within seconds, whereas the Bitcoin network, for example, experiences an approximately 10-minute delay between adding new blocks. If an onchain claim were issued on Bitcoin, it could take an hour or more before it is recognized by relying parties. Verifiers often need access to this blockchain data to compare revealed information against public hashes of that data or query an onchain revocation registry. The time delay for releasing blocks or for reading and processing newly published ones can thus affect the view that an application has of the current data.

**Chain Splits**:

Chain splits are another potential issue, such as that which occurred between Ethereum and Ethereum Classic. When certain kinds of disputes arise between users or stakeholders in a blockchain system, a single chain can split into two chains with a shared history up until the point of the split. If a smart contract existed on the chain prior to the split, it will have its state, history, and logic copied to both chains. This can cause confusion for users, especially during the time around the split. It may present further issues (e.g., replay attacks) such that a transaction that is valid on one chain is also valid on the other even if the transaction is only intended for a single chain. This may require relying parties and users to monitor both chains for some period of time.

**Blockchain Resiliency**:

As NISTIR 8202 states, "Traditional centralized systems are created and taken down constantly, and blockchain networks will likely not be different. However, because they are decentralized, there is a chance that when a blockchain network 'shuts down', it will never be fully shut down, and that there may always be some lingering blockchain nodes running. A defunct blockchain would not be suitable for a historical record, since without many publishing nodes, a malicious user could easily overpower the few publishing nodes left and redo and replace any number of blocks."

For an IDMS built on top of an underlying blockchain, it is important to carefully monitor the validators' activity and establish security thresholds and metrics to ensure that the increased risk of attacks on a declining blockchain are understood and considered acceptable. When a blockchain is deemed insecure, an identity management system may be migrated to a more secure one.

## 6.2   Introduction to Zero-Knowledge Protocols

Zero-knowledge protocols (ZK protocols, or ZKP) can play a fundamental role in blockchain-based identity management systems for transaction confidentiality, user identification, and presentation disclosure. Credentials can be taken as input to build presentations using zero-knowledge proofs, which allow subjects to control the amount of information disclosed to relying parties and the context in which the presentation takes place (see Section 4.3 on *Presentation Disclosure*).

The notion of zero-knowledge was first introduced in 1985 [63] and has since evolved into a class of algorithms with several practical applications [64, 65]. This section presents a high-level overview of zero-knowledge protocols and their role in identity management. Readers are encouraged to explore specialized publications, such as [66], and research work from initiatives, such as ZKProof.org [80], to gain a deeper understanding of zero-knowledge protocols.

**Definition and Properties**:

There are at least two parties in a ZK protocol: a *prover* and a *verifier*. The prover aims to convince the verifier that a statement is true without revealing any additional information. Below are common statement types, though the list is not exhaustive:

- An *equality statement* (e.g., the prover's bank account balance is equal to x) or non-equality statement
- An *inequality statement* (e.g., the prover's bank account balance exceeds x)
- A *range statement* (e.g., the prover's bank account balance is within interval [a, b]) or out-of-range statement
- A *membership statement* (e.g., the prover is on the client list of bank X) or non-membership statement
- A *hash pre-image statement* (e.g., the prover knows a secret value x that hashes to a known hash H)
- A *Merkle proof statement* (e.g., the prover knows a member of a Merkle tree with a known Merkle root R)

Generally, there are two kinds of ZK protocols: *interactive* and *non-interactive*. In an interactive ZK protocol, the prover and verifier engage in at least three rounds of communication exchange. Such protocols permit the verifier to submit challenges to the prover whereby the prover replies with responses that reinforce the validity of the prover's original statement. There is no challenge-response interaction in non-interactive ZK protocols, though there is sometimes a common reference string shared in advance by both parties.

A ZK protocol produces a proof that is sent to the verifier. For statement S, prover P, and verifier V, the resulting proof $\pi$ must satisfy the three following properties to be considered secure:

- *Completeness*: If S is true, then $\pi$ will convince V that S is true with overwhelming probability.
- *Soundness*: If S is false, then the probability that P can convince V that S is true is negligible.
- *Zero-knowledge*: If S is true, then V learns nothing from $\pi$ besides the fact that S is true.

The soundness property captures the inability of a prover P to convince the verifier V of a false assertion. If, for example, P can cheat with probability 1/3, then the ZK protocol may need to be repeated n times to reduce the soundness error from 1/3 to $1/3^n$.

The zero-knowledge property can be statistical or computational. If the verifier is assumed to have unlimited computational resources but learns no additional information from the protocol, then the protocol is considered to achieve statistical zero knowledge. If the zero-knowledge property holds by some assumption about the verifier's computational power, then the protocol achieves computational zero knowledge.

**Efficiency and Cost**:

The scalability and cost of ZK protocols depend on the succinctness of the proof. It measures the required storage size of the proof, proving time, and verification time. In a blockchain setting, the verification time is typically the most important consideration and proving time the least.

The trusted setup phase, required for some zero-knowledge protocols (e.g., the zk-SNARKs protocol implemented in Zcash [65]), involves a significant initial cost but then enables verifications of the proof to require fewer resources (i.e., it allows a statement to be proven many times by verifiers that have limited time and resources).

## 6.3   Presentation Sharing and Data Mining

This section discusses mechanisms at a high level (e.g., those based on zero-knowledge) to control the context in which presentations may be used by relying parties for data mining and data exchanges with third parties (i.e., other relying parties). Those mechanisms represent advanced research topics that could lead to the emergence of novel data broker business models and are mostly out of scope for this paper.

**Convincing Power**:

When a subject discloses a presentation to a relying party (as discussed in Section 4.3), information is revealed, and that cannot be undone. The relying party can then share that information with other relying parties. However, in some schemes, such as interactive zero-knowledge protocols, relying parties are, by design, unable to convince other relying parties that a statement—that they were themselves proven beforehand by a subject was true—is true. An interactive proof typically only convinces a single verifier that has established a direct and authenticated contact with the prover. In contrast, non-interactive protocols may convince multiple verifiers simultaneously and possibly at a later date.

Schemes also exist where ZK protocols allow for privacy-preserving querying of credential revocation registries (e.g., some forms of cryptographic accumulators).

**Benefits of Credential Properties**:

Presentations can take the form of credentials to benefit from the properties that credentials have. For instance, a presentation may have the ability to be accessed conditionally (see *User-Mintable, Predefined, Non-Fungible Token* in Section 4.4.2.1) and to be transferable (see *Non-Fungible Token Registry* in Section 4.4.2.1).

Such presentations can also be used to derive a limited number of presentations, like in the scheme described in [67].

**Presentation Encapsulation**:

A relying party may receive and encapsulate presentations into other presentations. In that case, the issuer of the encapsulated presentation is not the issuer of the original presentation.

However, it allows the relying party to verify a snapshot of a presentation for another relying party (a timestamp and signature may be added).[10]

## 6.4 Ecosystem Convergence

This section provides a few high-level considerations about wallet interoperability and cross-ledger integration, the development of which is a key ecosystem catalyzer for blockchain-based identity management. Contingent upon this is the identification of criteria and best practices to understand which architecture designs are relevant, depending on the use cases at stake, and how to assemble them into suitable solutions. Although this paper does not provide such recommendations, it notes that they could help inform decisions on how to use an existing system as a service, integrate one to a given solution, or build and deploy a new one.

**Interoperable Wallets**:

Standards such as BIP-32 and ERC-20 facilitated the emergence of interoperable wallets for cryptocurrencies. Similarly, the emerging standards for identity management discussed in this paper could lead to interoperable, user-controlled wallets that store and manage identifiers and credentials for multiple blockchain-based IDMSs alongside cryptocurrencies, cryptocollectibles, and other types of blockchain-based digital assets. Password managers could also be integrated for the storage and management of traditional identifiers and credentials. Such interoperable wallets could serve as gateways controlled by users who would thus authenticate to third-party applications and services from a common interface.

Standards and techniques for advanced security and private key custody schemes are also crucial, such as those that rely on threshold cryptography. For example, ZenGo [64] has developed a wallet that uses threshold signatures to create two secret shares that take the role of a user's private key when they are combined (this may be extended with more shares and trusted third parties). Secret shares are also used in the Horcrux protocol [69],[11] which follows the Biometric Open Protocol Standard [70][12] to power blockchain-based authentication with biometric information.

---

[10] Since non-interactive zero-knowledge (NIZK) protocols are potentially transferrable, the original verifier could claim the original NIZK proof as their own while interacting with third-party verifiers. [68] provides a way to tie NIZK proofs to the identity of the original prover such that when the original verifier presents it to a third party, the third party will understand that it was the original prover and not the original verifier who issued the original proof.

[11] In this protocol, biometric data is collected by a device owned by the user then divided into multiple shares. One of these shares is sent to a dedicated server, which selects a blockchain, creates a DID for the biometric share and stores the resulting DID document using offchain storage providers. The other biometric shares can similarly be assigned to other blockchains, creating more DIDs. As a result, the original biometric data can act as a junction between different identity management platforms. This can help create more robust, blockchain-agnostic solutions.

[12] The Biometric Open Protocol Standard (BOPS) was introduced by IEEE under reference 2410-2018. It provides a framework for supporting biometric authentication methods. This standard also offers guidance for identification, access control, and auditing capabilities. Dedicated API designs, device requirements, and security and privacy considerations are also introduced.

**Cross-Ledger Integration**:

There are several ways blockchain-based identity management systems can integrate with one another and/or be part of a common larger structure, such as:

- *Universal resolver*: As mentioned in Section 3.3 on *Emerging Standards*, the blockchain-agnostic Universal DID Resolver maintained by the DIF allows the integration of any identity management system, which can then be queried by the users through a common interface.
- *Bridges*: The capabilities of a given system may be integrated into another system by implementing the libraries provided by the former system in the form of onchain logic (e.g., smart contracts) in the latter system. For example, Cordentity [71] is a Corda smart contract that integrates Hyperledger Indy capabilities into the Corda platform. Thus, Corda ledger transactions can be contingent on credentials managed with a Hyperlerdger Indy-based blockchain. Additionally, SecureKey has explored integrating Hyperledger Indy capabilities in Verified.me, its Hyperledger Fabric-based identity management system [72].

## 7 Use Cases

Blockchain-based identity could become a fundamental architectural component of tomorrow's web, bringing forward novel applications and services. Below is a brief discussion on simple use cases using some of the terminology and architectural considerations introduced in this paper.

Applications can be public-facing (for sharing publicly verifiable data) or privacy-preserving (to provide solutions for individual users) in industries such as:

- Banking: account opening, fraud prevention, compliance with anti-money laundering (AML) and KYC laws, proof of funds, credit risk evaluation, as well as ownership, exchange, and trading of financial assets
- Data brokerage: exchanges of datasets
- Supply chain management: asset traceability and lifecycle, networks of sensors, and the Internet of Things
- Education: issuance of transcripts, diplomas, and certifications that can then serve as verifiable credentials in job applications
- Healthcare: issuance of prescriptions, submission of claims to health insurance, and sharing of health records
- Government services: issuance of driver's licenses and birth certificates, maintenance of public registries of voters
- Public safety: management of sets of equipment and reliable communication permissions
- Manufacturing: representation of ownership of 3D models
- Transportation: identification of autonomous vehicles

Below are three use cases that may further bolster an understanding of blockchain-based identity management.

**Renting a Vehicle**:

This use case considers an individual who is proving to a car rental company that they meet all the requested requirements without disclosing more information than what is strictly needed.

This takes place through a system that enables the individual to build and disclose the necessary proofs derived from pre-existing credentials that they own on a given IDMS. Those credentials are self-contained—existing as digital files stored in the individual's wallet application—with an onchain revocation registry. They include an unexpired and valid driver's license, an insurance certificate (showing that the individual has sufficient coverage), and a bank statement (showing that the individual has the means to pay the deductible in case of an accident). Rather than sharing the credentials in their entirety to the rental company, the presentation built by the system allows the individual to combine the derived information from each credential (as shown in Figure 9) and prove that they meet all the requirements. It may not even be necessary to disclose the full name of the individual.

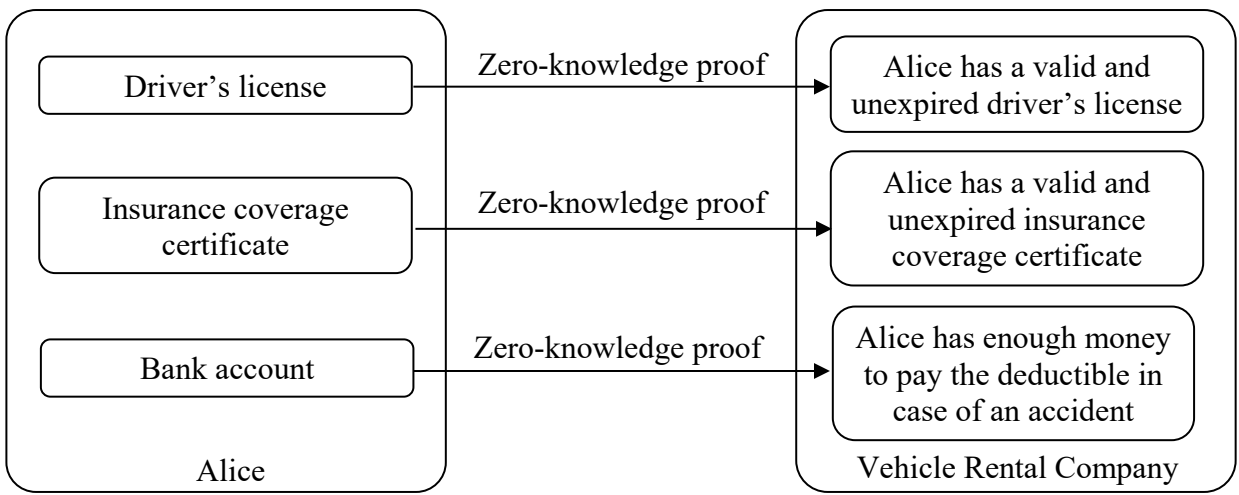

**Figure 9: Minimal Disclosure to Rent a Car**

An alternative version of this scenario is that of an employee who is renting a vehicle on behalf of the company that they work for. In this case, the company can delegate access to some of its credentials to the employee so that information derived from these credentials can then be added to the presentation the employee discloses to the rental car company.

More generally, presentation disclosure can involve any type of entity, including vehicles, which can have their own blockchain-based identifiers as shown in the Vehicle Identity (VID) standard proposed by the Mobility Open Blockchain Initiative (MOBI) [73] for vehicle-to-vehicle and vehicle-to-infrastructure communications and payments systems (e.g., automated traffic routing, toll payments, parking services, maintenance records).

**Exchanging Event Tickets and Coupons**:

This use case considers a system to issue and manage transferable event tickets and coupons (e.g., for concerts, conferences).

The system is owned by a ticketing company that controls initial identity proofing and user registration. Once registered, event organizers can issue transferable tickets to registered users (in the form of non-fungible tokens). Although the initial registration is controlled by the system owner, users can transfer tickets on their own without any further approval from the system owner. For instance, a ticket owner may be able to exchange the ticket for one on another date, give it to a friend, or sell it. After attending a concert, an individual may keep the digital ticket as a souvenir and add an attestation of it on social media to connect with other attendees and artists.

The system also implements a loyalty program to earn rewards and attend other events. It periodically distributes redeemable coupons (in the form of a Merkle airdrop of non-fungible tokens) to the customer base that can then be used to claim discounts to attend new events. While these coupons have an expiration date and were issued to a certain group of individuals, they too can be made transferable. For example, an individual that receives a coupon but prefers to give it to a friend can do so, thus allowing the event organizers to reach a wider target audience.

**Verifying Business Identity**:

This use case considers a system to issue and manage public identifiers and credentials for businesses.

It could follow either a top-down authority model, such as a system owned by some central authority (e.g., government, chamber of commerce) that identity proofs businesses, or a bottom-up authority model (e.g., applications that self-register in an application store).

Once originated in the system, businesses have identifiers and can be issued credentials for certificates, accreditations, and other types of verifiable information that will serve for authentication during commercial interactions. Additionally, businesses can become credential issuers with respect to their staff members. In turn, staff members can build verifiable presentations proving that they act on behalf of their company.

From an architectural standpoint, identifiers and credentials are both stored in a common onchain registry, allowing for any participant to publicly verify the information of a given business.

The solutions proposed by the Global Legal Entity Identifier Foundation, uPort [74], and VON [55] illustrate this use case.

## 8      Conclusion

Blockchain-based identity management is an emerging field that holds great promise in providing an alternative to centralized user identity data custody usually found in traditional and federated models. This paper is meant to provide the reader with a general understanding of the benefits, challenges, and opportunities of such systems, both for users and businesses. It discusses foundational building blocks of blockchain-based identity management and current standardization efforts. It then identifies emerging architectural designs and their properties. Finally, the paper reviews security and risk management issues and gives blockchain-specific considerations before concluding with some example use cases.

The paper highlights the potential for blockchain-based identity management systems—from which novel data ownership and governance models can be developed—to make it easier for users to control the custody of their own identifiers and credentials. Capabilities for selective, verifiable data sharing between users and relying parties can be built into these systems, reducing the dependence on trusted intermediaries.

There are still scalability, security, and privacy considerations that must be carefully scrutinized to build viable digital identity solutions using blockchains, zero-knowledge proofs, second layer protocols, and related technologies. That said, if properly addressed, blockchain-based identity could become a fundamental architectural component of tomorrow's web.

## Appendix A—Acronyms

Selected acronyms and abbreviations used in this paper are defined below.

| | |
|---|---|
| ACT-IAC | American Council for Technology and Industry Advisory Council |
| AML | Anti-Money Laundering |
| API | Application Programming Interface |
| BIP | Bitcoin Improvement Proposal |
| DAO | Decentralized Autonomous Organization |
| DID | Decentralized Identifier |
| DIF | Decentralized Identity Foundation |
| DLT | Distributed Ledger Technology |
| DNS | Domain Name System |
| ERC | Ethereum Request for Comments |
| ETH | Ethereum |
| FIM | Federated Identity Management |
| HD | Hierarchical Deterministic |
| HTTP | Hyper-Text Transfer Protocol |
| ICO | Initial Coin Offering |
| IDMS | Identity Management System |
| IEEE | Institute of Electrical and Electronics Engineers |
| IETF | Internet Engineering Task Force |
| IP | Internet Protocol |
| IPFS | Inter-Planetary File System |
| ISO | International Organization for Standardization |
| ITL | Information Technology Laboratory |
| JSON | JavaScript Object Notation |
| JSON-LD | JavaScript Object Notation for Linked Data |
| JWT | JSON Web Token |
| KYC | Know Your Customer |
| MOBI | Mobility Open Blockchain Initiative |
| NFC | Near-Field Communication |
| NFT | Non-Fungible Token |

| NIST | National Institute of Standards and Technology |
| NISTIR | National Institute of Standards and Technology Internal Report |
| NIST SP | National Institute of Standards and Technology Special Publication |
| PII | Personally-Identifiable Information |
| QR | Quick Response |
| RBFT | Redundant Byzantine Fault Tolerance |
| RFC | Request for Comments |
| RWOT | Rebooting the Web-of-Trust |
| SAML | Security Assertion Markup Language |
| SDK | Software Development Kit |
| SSI | Self-Sovereign Identity |
| SSO | Single Sign-On |
| TLS | Transport Layer Security |
| UI | User Interface |
| URI | Uniform Resource Identifier |
| URL | Uniform Resource Locator |
| UTXO | Unspent Transaction Output |
| VID | Vehicle Identity |
| VON | Verifiable Organizations Network |
| W3C | World Wide Web Consortium |
| XDI | eXtensible Data Interchange |
| ZK | Zero-Knowledge |
| ZKP | Zero-Knowledge Protocol |

## Appendix B—Glossary

| | |
|---|---|
| Airdrop | A distribution of digital tokens to a list of blockchain addresses. |
| Asymmetric-Key Cryptography [11] | A cryptographic system where users have a private key that is kept secret and used to generate a public key (which is freely provided to others). Users can digitally sign data with their private key and the resulting signature can be verified by anyone using the corresponding public key. Also known as Public-key cryptography. |
| Authentication [75] | Verifying the identity of a user, process, or device, often as a prerequisite to allowing access to resources in an information system. |
| Blockchain [11] | Blockchains are distributed digital ledgers of cryptographically signed transactions that are grouped into blocks. Each block is cryptographically linked to the previous one (making it tamper evident) after validation and undergoing a consensus decision. As new blocks are added, older blocks become more difficult to modify (creating tamper resistance). New blocks are replicated across copies of the ledger within the network, and any conflicts are resolved automatically using established rules. |
| Claim | A characteristic or statement about a subject made by an issuer as part of a credential. |
| Consensus Model [11] | A process to achieve agreement within a distributed system on the valid state. <br><br> Also known as a *consensus algorithm*, *consensus mechanism*, *consensus method*. |
| Credential | A set of one or more claims made by an issuer. A credential is associated with an identifier. |
| Cryptocurrency [11] | A digital asset/credit/unit within the system, which is cryptographically sent from one blockchain network user to another. In the case of cryptocurrency creation (such as the reward for mining), the publishing node includes a transaction sending the newly created cryptocurrency to one or more blockchain network users. <br><br> These assets are transferred from one user to another by using digital signatures with asymmetric-key pairs. |
| Cryptographic Hash Function [11] | A function that maps a bit string of arbitrary length to a fixed-length bit string. Approved hash functions satisfy the following properties: |

1.  (*Preimage resistant*) It is computationally infeasible to compute the correct input value given some output value (the hash function is "one way").
2.  (*Second preimage resistant*) One cannot find an input that hashes to a specific output.
3.  (*Collision resistant*) It is computationally infeasible to find any two distinct inputs that map to the same output.

| | |
|---|---|
| Curation Market | A token-based organization model that aims at incentivizing and coordinating market participants around the curation of some information. *Term introduced by Simon de la Rouviere.* |
| Custodian | An entity acting on behalf of another entity with respect to their identifiers and/or credentials. |
| Decentralized Autonomous Organization | A system that is not controlled by a single entity or leader, and that, instead, uses onchain registries and logic to establish some form of self-sustainable organizational structure (e.g., through market incentives, network effects, and protocol designs). |
| Digital Token | A representation of a particular asset that typically relies on a blockchain or other types of distributed ledgers. Also known as Token. |
| Entity | A person, organization, or thing. |
| Factory Smart Contract | A smart contract that creates, and sometimes, manages other smart contracts. |
| Hash [11, Adapted] | The output of a hash function (e.g., hash(data) = digest). Also known as a message digest, digest, hash digest, or hash value. |
| Holder | A custodian holding a credential on behalf of an entity. |
| Identifier | A blockchain address or other pseudonym that is associated to an entity. |
| Issuer | An entity that issues a credential about a subject on behalf of a requester. |
| JSON Web Token [76, Adapted] | A JSON Web Token (JWT) is a data exchanged format comprised of a header, a payload, and a signature where the header and the payload take the form of JSON objects. They are encoded and concatenated with the aggregate being signed to generate a signature. |
| Linked Data | A method for interconnecting data structures to promote interpretability. *Term introduced by Tim Berners-Lee.* |

| Merkle Airdrop | A scheme to distribute the entitlement to redeem a digital token to a list of blockchain addresses in a single transaction rather than distributing the tokens themselves in a batch of transactions as in a standard airdrop. The list must be available to the participants so that they can build the proof needed to redeem the token, called Merkle proof (as it relies on a Merkle tree). |
| --- | --- |
| Merkle Tree [11] | A data structure where the data is hashed and combined until there is a singular root hash that represents the entire structure. |
| Mintable | Refers to the ability of a digital token to be created. |
| Node [11] | An individual system within the blockchain network. |
| Non-Fungible | Refers to something that is uniquely identifiable (i.e., not replaceable / interchangeable). |
| Offchain | Refers to data that is stored, or a process that is implemented and executed, outside of any blockchain system. |
| Onchain | Refers to data that is stored, or a process that is implemented and executed, within a blockchain system. |
| Oracle | A source of data from outside a blockchain that serves as input for a smart contract. |
| Pairwise-Pseudonymous Identifier | An identifier solely used for interactions with a particular relying party. |
| Presentation | Information derived from one or more credentials that a subject discloses to a verifier to communicate some quality about a subject. |
| Relying Party | An entity that receives information about a subject from a verifier. |
| Requester | An entity that makes a request to an issuer to issue a credential about a subject. |
| Resolver | Software that retrieves data associated with some identifier. |
| Selective Disclosure Mechanism | A scheme for presentation disclosure that allows to granularly control the amount of information that is being released. |
| Single-Use Identifier | An identifier discarded immediately after being used for one particular interaction. |
| Smart Contract [11] | A collection of code and data (sometimes referred to as functions and state) that is deployed using cryptographically signed transactions on |

|  | the blockchain network. The smart contract is executed by nodes within the blockchain network; all nodes must derive the same results for the execution, and the results of execution are recorded on the blockchain. |
| --- | --- |
| Subject | An entity that receives one or more credentials from an issuer. |
| System Owner | An entity that owns a given identity management system. |
| Unlinkability | The extent to which a relying party is unable to link a given identifier to other ones a subject may own. |
| Uniform Resource Identifier [77] | A compact sequence of characters that identifies an abstract or physical resource available on the Internet. |
| Verifier | An entity that verifies the validity of a presentation on behalf of a relying party. |
| Wallet [11] | Software used to store and manage asymmetric-keys and addresses used for transactions. |
| Zero-Knowledge Proof | A cryptographic scheme where a prover is able to convince a verifier that a statement is true, without providing any more information than that single bit (that is, that the statement is true rather than false). |